\documentclass[fleqn]{article}
\usepackage{tikz-feynman}
\usepackage{cite}
\usepackage{mhchem}

\usepackage{slashed}
\usepackage[pdftex]{hyperref}
\usepackage{amsmath, amsthm, amssymb,slashed,mathrsfs}
\parskip=\baselineskip

\font\teneurm=eurm10 \font\seveneurm=eurm7 \font\fiveeurm=eurm5
\newfam\eurmfam
\textfont\eurmfam=\teneurm \scriptfont\eurmfam=\seveneurm
\scriptscriptfont\eurmfam=\fiveeurm

 \font\teneusm=eusm10 \font\seveneusm=eusm7 \font\fiveeusm=eusm5
\newfam\eusmfam
\textfont\eusmfam=\teneusm \scriptfont\eusmfam=\seveneusm
\scriptscriptfont\eusmfam=\fiveeusm

\font\tencmmib=cmmib10 \skewchar\tencmmib='177
\font\sevencmmib=cmmib7 \skewchar\sevencmmib='177
\font\fivecmmib=cmmib5 \skewchar\fivecmmib='177
\newfam\cmmibfam
\textfont\cmmibfam=\tencmmib \scriptfont\cmmibfam=\sevencmmib
\scriptscriptfont\cmmibfam=\fivecmmib

\parskip=\baselineskip

\def\p{\mbox{\boldmath$\displaystyle\mathbf{p}$}}

\def\0{\mbox{\boldmath$\displaystyle\mathbf{0}$}}

\def\x{\mbox{\boldmath$\displaystyle\mathbf{x}$}}

\begin{document}

\unitlength=1mm
%\begin{titlepage}
%\begin{flushright}
%CERN-TH-PH/2009-019\\
%\end{flushright}
%begin{flushright}
%hep-th/yymm.nnnn
%\end{flushright}
\vskip 1.5in
\begin{center}
{\bf\Large{Interactions and oscillations of coherent flavor eigenstates\\ in beta decay}}\vskip 0.5cm
{Cheng-Yang Lee} \vskip 0.3in 
{
\small{\textit{Manipal Center for Natural Sciences, Center of Excellence,\\ Manipal Academy of Higher Education,\\ Dr. T.~M.~A.~Pai Planetarium Building, Manipal 576104, Karnataka, India}}}

{
\small{\textit{Center for Theoretical Physics, College of Physical Science and Technology,\\
Sichuan University, Chengdu, 610064, China\\
Email: cylee@scu.edu.cn}}}

\begin{abstract}
The theory pioneered by Blasone and Vitiello describes massive neutrinos as generalized coherent flavor eigenstates within the extended Standard Model. In this paper, we compute the neutron $\beta$ decay spectrum for the Blasone-Vitiello theory in a model with two neutrinos. For relativistic neutrinos, the obtained spectrum is in agreement with the Standard Model. However, there are discrepancies when the kinetic energy of the electron is close to the end point energy. The possibility of measuring the discrepancies in future experiments is discussed.
\end{abstract}

\end{center}
\vskip 0.5in

%------------------------------------------
\section{Introduction}
%------------------------------------------

Massive neutrinos and their flavor oscillations are direct evidences for physics beyond the Standard Model (SM)~\cite{Cleveland:1998nv,Fukuda:1998mi,Aguilar:2001ty,Ahmad:2002jz,
Aguilar:2007it,Adamson:2011qu,Abe:2011fz,An:2012eh,Ahn:2012nd}. 
%For relativistic neutrinos, the Pontecorvo theory provides an adequate phenomenological description of flavor oscillations where a neutrino state $|\nu_{\alpha}\rangle$ is given by a linear combination of mass eigenstate $|\p_{i},m_{i}\rangle$ of mass $m_{i}$ and momentum $\p_{i}$:
%\begin{equation}
%|\nu_{\alpha}\rangle\equiv\sum_{i}U^{*}_{\alpha i}|p_{i},m_{i}\rangle \label{eq:PS}
%\end{equation}
%with $U_{\alpha i}$ being the unitary mixing matrix. Despite its success, from a theoretical perspective, the Pontecorvo theory is incomplete because it is based on quantum mechanics (QM) and not quantum field theory (QFT). In fact, it was shown in ref.~\cite{Giunti:1991cb} that eq.~(\ref{eq:PS}) is incompatible with the extended SM except in the relativistic limit or for states of degenerate masses.\footnote{In ref.~\cite{Giunti:1991cb}, neutrinos are Majorana fermions but the results can be generalized to Dirac neutrinos. In this  paper, we take neutrinos to be of the Dirac type.} 
Based on the extended SM with massive neutrinos, the works of Shrock~\cite{Shrock:1980vy,Shrock:1980ct,Shrock:1981wq} and Giunti, Kim and Lee~\cite{Giunti:1991cb,Giunti:1993se} provided a unified description of neutrino interactions and oscillations. In these works, the quantum field associated with a Dirac neutrino of flavor $\alpha$ is given by
\begin{equation}
\nu_{\alpha}(x)=\sum_{i}U_{\alpha i}\nu_{i}(x) \label{eq:ff}
\end{equation}
where $\nu_{i}(x)$ is a Dirac field of mass $m_{i}$ and they take part in the weak interactions as mass eigenstates. In the case of neutron $\beta$ decay to be studied in sec.~\ref{exp}, the flavor-conserving process $n\rightarrow p+e^{-}+\overline{\nu}_{e}$ for massless neutrinos is replaced by $n\rightarrow p+e^{-}+\overline{\nu}_{i}$ where $i=1,\cdots, N$ and $N$ is the number of mass eigenstates. While mass eigenstates do not oscillate, their associated physical processes are flavor-violating. As a result, flavor oscillations can be explained when the production and detection of mass eigenstates are taken into account~\cite{Giunti:1993se}.

The above works provide an adequate description of neutrinos capable of explaining the observed phenomena but it is not the only possibility. The theory pioneered by Blasone and Vitiello (BV) provides an alternative description by uncovering new mathematical structures in eq.~(\ref{eq:ff}) that allows neutrinos to be described as generalized coherent flavor eigenstates~\cite{Blasone:1995zc}. Since neutrinos present the most concrete evidence for physics beyond the SM, it is important and worthwhile to fully explore all possibilities presented to us by the theory.

An important task for the BV theory is to study neutrino interactions. One such calculation was carried out in refs.~\cite{Li:2006qt,Blasone:2006jx} but was incomplete. The main concern for the theory is that since flavor eigenstates undergo oscillations, it is not obvious how their asymptotic free states far before and after interactions which are needed to compute the $S$-matrix are defined. In this paper, we compute the neutron $\beta$ decay spectrum  for the BV theory in a model with two neutrinos. For relativistic neutrinos, the spectrum is agreement with the SM.
%~\footnote{In this paper, the (extended) Standard Model strictly refers to a model with three (massive) massless neutrinos. From here onwards, unless otherwise specified, we work with the BV and Fermi theory of two neutrinos}. 
However, there is an important conceptual difference between the BV theory and the extended SM. In the BV theory, the neutron $\beta$ decay has two channels $n\rightarrow p^{+}+e^{-}+\overline{\nu}_{e}$ and $n\rightarrow p^{+}+e^{-}+\overline{\nu}_{\mu}$. In the short-time approximation just before and after the interaction, the finite-time $S$-matrix for the flavor-violating process vanishes. In the limit far before and after the interaction ($t\rightarrow\pm\infty$), the flavor-violating process is non-vanishing. This can be explained by the asymptotic behavior of oscillations which is naturally incorporated in the BV theory. For relativistic neutrinos, the spectrum for each flavor factorizes to the SM spectrum multiplied by the average oscillation probability. Physically, this means that the detector is placed at a distance greater than the coherent oscillation length from the source. As the kinetic energy of the electron approaches the end point energy, discrepancies between the BV theory and the SM become significant. In the case of Tritium $\beta$ decay, the difference can be as large as $7\%$.

The paper is organized as follows. In sec.~\ref{coherent}, we review the BV theory. In sec.~\ref{exp}, we compute the neutron $\beta$ decay spectrum using the BV theory. Finally, we conclude by discussing the relevance of our results to the KATRIN~\cite{Osipowicz:2001sq,WEINHEIMER2002141} and the upcoming PTOLEMY experiment~\cite{Betts:2013uya}.

%-------------------------------------------------------------------------
\section{Generalized coherent flavor eigenstates}\label{coherent}
%-------------------------------------------------------------------------

%In this section, we present a consistent QFT description of neutrinos as flavor eigenstates using the BV theory~\cite{Blasone:1995zc,Blasone:1998hf}.

The BV theory is based on $\nu_{\alpha}(x)$ given by eq.~(\ref{eq:ff}).
%\begin{equation}
%\nu_{\alpha}(x)=\sum_{i}U_{\alpha i}\nu_{i}(x)\label{eq:ff}
%\end{equation}
%where $\nu_{i}(x)$ is a Dirac field of mass $m_{i}$.  
%The no-go theorems proved in ref.~\cite{Giunti:1991cb} and the previous section have shown the latter to be incompatible with QFT. While these problems are absent for $\nu_{\alpha}(x)$, because it is a linear combination of Dirac fields of different masses, it is not obvious how $\nu_{\alpha}(x)$ can be expanded in terms of flavor annihilation and creation operators. 
While $\nu_{\alpha}(x)$ is Poincar\'{e}-covariant, it is not obvious how $\nu_{\alpha}(x)$ can be expanded in terms of flavor annihilation and creation operators that satisfy the canonical anti-commutation relations. To the best of our knowledge, this problem was initially solved by Chang et al.~\cite{Chang:1980qw} and later studied in more depth by BV. It was shown that there exists an expansion of $\nu_{\alpha}(x)$ in terms of flavor operators for the theory of two and three neutrinos that satisfy the canonical anti-commutation relations~\cite{Blasone:1995zc,Blasone:2002jv}. This was later generalized to arbitrary number of neutrinos~\cite{Hannabuss:2000hy,Ji:2002tx}. As a result, neutrinos can be described by the generalized coherent flavor eigenstates~\footnote{From now onwards, we will simply refer to the generalized coherent flavor eigenstates as flavor eigenstates.} .

We now review the solution to the problem of expanding $\nu_{\alpha}(x)$ in terms of flavor operators as presented in refs.~\cite{Blasone:1995zc,Blasone:2002jv}. For simplicity, we consider the theory of two neutrinos where
\begin{eqnarray}
&& \nu_{e}(x)=+\cos\theta\,\nu_{1}(x)+\sin\theta\,\nu_{2}(x),\label{eq:cs1} \\
&& \nu_{\mu}(x)=-\sin\theta\,\nu_{1}(x)+\cos\theta\,\nu_{2}(x).\label{eq:cs2}
\end{eqnarray}
Comparing to the theory of three neutrinos, apart from the absence of $\mathsf{CP}$ violation, the oscillation formulae are simpler but still contain the essential features.  The crucial observation was that eqs.~(\ref{eq:cs1}) and (\ref{eq:cs2}) are Bogoliubov transformations. They can be rewritten in the form
\begin{eqnarray}
&& \nu_{e}(x)=G^{-1}_{\theta}(t)\nu_{1}(x)G_{\theta}(t),\label{eq:g1}\\
&& \nu_{\mu}(x)=G^{-1}_{\theta}(t)\nu_{2}(x)G_{\theta}(t)\label{eq:g2}
\end{eqnarray}
where $G_{\theta}(t)$ is a unitary operator defined as
\begin{equation}
G_{\theta}(t)\equiv\exp\left[\theta\int d^{3}x\left(\nu^{\dag}_{1}(x)\nu_{2}(x)-\nu^{\dag}_{2}(x)\nu_{1}(x)\right)\right].
\end{equation}

The field $\nu_{i}(x)$ takes the form
\begin{eqnarray}
\nu_{i}(x)&=&(2\pi)^{-3/2}\int\frac{d^{3}p}{\sqrt{2E_{i}}}\sum_{\sigma}
\Big{[}e^{i\mathbf{p\cdot x}}u_{i}(\p,\sigma)a_{i}(\p,\sigma,t)
+e^{-i\mathbf{p\cdot x}}v_{i}(\p,\sigma)b_{i}^{\dag}(\p,\sigma,t)\Big{]} \label{eq:df}
\end{eqnarray}
for $i=1,2$ where $p^{\mu}_{i}=(E_{i},\p)$ and $E_{i}=\sqrt{|\p|^{2}+m^{2}_{i}}$. The operators at time $t$ are given by
\begin{equation}
a_{i}(\p,\sigma,t)=e^{-iE_{i}t}a(\p,\sigma),\quad b_{i}(\p,\sigma,t)=e^{-iE_{i}t}b(\p,\sigma)
\end{equation}
where $a_{i}(\p,\sigma)$ and $b_{i}(\p,\sigma)$ satisfy the standard canonical anti-commutation relations
\begin{eqnarray}
\{a_{i}(\p,\sigma),a_{j}^{\dag}(\p',\sigma')\}&=&\{b_{i}(\p,\sigma),b_{j}^{\dag}(\p',\sigma')\}\nonumber\\
&=&\delta_{\sigma\sigma'}\delta_{ij}\delta(\p-\p').
\end{eqnarray}
The vacuum state for the theory is defined as a tensor product of the vacuum for the two mass eigenstates $|\,\,\rangle\equiv|\,\,\rangle_{1}\otimes|\,\,\rangle_{2}$ so it is annihilated by $a_{i}(\p,\sigma)$ and $b_{i}(\p,\sigma)$. The particle and anti-particle mass eigenstates are given by
\begin{equation}
a_{i}^{\dag}(\p,\sigma)|\,\,\rangle=|p,m_{i},\sigma\rangle,\hspace{0.5cm}
b_{i}^{\dag}(\p,\sigma)|\,\,\rangle=|\overline{p},m_{i},\sigma\rangle.
\end{equation}

Substituting eq.~(\ref{eq:df}) into (\ref{eq:g1}) and (\ref{eq:g2}), we may expand the flavor field as
\begin{eqnarray}
\nu_{\alpha}(x)&=&(2\pi)^{-3/2}\int\frac{d^{3}p}{\sqrt{2E_{i}}}\sum_{\sigma}
\Big{[}e^{i\mathbf{p\cdot x}}u_{i}(\p,\sigma)a_{\alpha}(\p,\sigma,t)
+e^{-i\mathbf{p\cdot x}}v_{i}(\p,\sigma)b^{\dag}_{\alpha}(\p,\sigma,t)\Big{]} \label{eq:ffa}
\end{eqnarray}
where the flavor operators at time $t$ are defined as
\begin{eqnarray}
&&a_{\alpha}(\p,\sigma,t)\equiv G^{-1}_{\theta}(t)[e^{-iE_{i}t}a_{i}(\p,\sigma)]G_{\theta}(t),\label{eq:aa}\\
&&b_{\alpha}(\p,\sigma,t)\equiv G^{-1}_{\theta}(t)[e^{-iE_{i}t}b_{i}(\p,\sigma)]G_{\theta}(t).
\label{eq:bb}
\end{eqnarray}
For eqs.~(\ref{eq:ffa}-\ref{eq:bb}), when $\alpha=e$, $i=1$ and for $\alpha=\mu$, $i=2$. The flavor operators annihilate the time-dependent flavor vacuum 
\begin{equation}
|\,\,\rangle_{\theta,t}\equiv G^{-1}_{\theta}(t)|\,\,\rangle\label{eq:fv}
\end{equation}
and their Hermitian conjugations create single neutrino and anti-neutrino states
\begin{eqnarray}
&& a^{\dag}_{\alpha}(\p,\sigma,t)|\,\,\rangle_{\theta,t}=|\p,\alpha,\sigma,t\rangle,
\label{eq:fs1}\\
&& b^{\dag}_{\alpha}(\p,\sigma,t)|\,\,\rangle_{\theta,t}=|\bar{\p},\alpha,\sigma,t\rangle.
\label{eq:fs2}
\end{eqnarray}

The states given in eqs.~(\ref{eq:fs1}) and (\ref{eq:fs2}) are identified to be (coherent) flavor eigenstates. The name is derived from the fact that they are eigenstates to the charge operator
\begin{equation}
{Q}_{\alpha}=\int d^{3}p\sum_{\sigma}\left[a^{\dag}_{\alpha}(\p,\sigma,t)a_{\alpha}(\p,\sigma,t)-
b^{\dag}_{\alpha}(\p,\sigma,t)b_{\alpha}(\p,\sigma,t)\right],\quad{\alpha=e,\mu}
\end{equation}
that does not commute with the Hamiltonian. Therefore, charges of $Q_{\alpha}(t)$ are not conserved so the flavor eigenstates must necessarily undergo oscillations as they evolve in time. In ref.~\cite{Blasone:2003eh}, the temporal evolution of the flavor eigenstates are quantified by computing the expectation values $\langle\p,\alpha,\sigma|Q_{\alpha}(t)|\p,\alpha,\sigma\rangle$  which is interpreted as the oscillation probability. This interpretation is elegant and seems reasonable. However, it is important to realize that the charge operator $Q_{\alpha}(t)$, despite being Hermitian, is not an observable that is measured in experiments. In real experiments, neutrino flavors are not directly measured. Instead, flavor oscillations are inferred by measuring the charged leptons in the initial and final states. Therefore, one approach to study the physical properties of the flavor eigenstates is to consider interactions where the effects of production and detection separated by finite time and distance are taken into account. Such an analysis has been performed in ref.~\cite{Giunti:1993se} for neutrino mass eigenstates but we will not consider it here. In this paper, we compute the spectrum of neutron $\beta$ decay for the flavor eigenstates. We show that for relativistic neutrinos, the spectrum factorizes to the SM spectrum multiplied by the standard oscillation probabilities. This shows that our identification of states given eqs.~(\ref{eq:fs1}) and (\ref{eq:fs2}) as flavor eigenstates is appropriate.

It is important to note that in eq.~(\ref{eq:ffa}), we chose to expand $\nu_{\alpha}(x)$ in a particular basis where the Dirac spinors are of masses $m_{e}=m_{1}$ and $m_{\mu}=m_{2}$. But this choice is not unique, the field $\nu_{\alpha}(x)$ can also be expanded using Dirac spinors of arbitrary mass parameter $\widetilde{m}_{\alpha}$ thus calling into question on whether the flavor eigenstates constructed by BV are physical~\cite{Fujii:1998xa,Fujii:2001zv,Giunti:2003dg}. This concern is a reflection of the general fact that asymptotic free quantum fields have infinitely many in-equivalent vacua but only the vacuum that corresponds to physically observed states is relevant~\cite{Blasone:2005ae}. In the case of neutrinos, the physically relevant mass eigenstates for $\nu_{i}(x)$ are of masses $m_{i}$ so we make the choice $m_{e}=m_{1}$ and $m_{\mu}=m_{2}$. However, it should be noted that if one interprets neutrino mixing as interactions with a constant non-Abelian gauge field, then a difference choice of masses have to be made~\cite{Blasone:2010zn}.

Equations~(\ref{eq:cs1}), (\ref{eq:cs2}) and (\ref{eq:ffa}) provide two different expansions for the flavor fields. By equating them and using the orthonormal relations between the Dirac spinors,  $a_{\alpha}(\p,\sigma,t)$ and $b_{\alpha}(\p,\sigma,t)$ can be expanded in terms of $a_{i}(\p,\sigma)$ and $b_{i}(\p,\sigma)$ (see app.~\ref{A}). 

One of the important features of the BV theory is the in-equivalence of states in the infinite volume limit. Specifically, the flavor and mass eigenstates are orthogonal and the flavor eigenstates at different times are also orthogonal:
\begin{equation}
\begin{array}{ll}
_{\theta,t}\langle\,\,|\,\,\rangle=0, &
\langle \p,m_{i},\sigma'|\p,\alpha,\sigma\rangle=0,\\
 _{\theta,t'}\langle\,\,|\,\,\rangle_{\theta,t}=0, &
\langle \p,\alpha,\sigma',t'|\p,\alpha,\sigma,t\rangle=0.\end{array}
\end{equation}
Details of the analysis can be found in refs.~\cite{Blasone:1995zc,Blasone:2005ae}. Another important feature is the non-trivial time evolution of the flavor operators.  Since the flavor eigenstates are not energy eigenstates, their time evolutions are more complicated. The explicit expressions of $a_{\alpha}(\p,\sigma,t)$ and $b_{\alpha}(\p,\sigma,t)$ in terms of $a_{\alpha}(\p,\sigma)$ and $b_{\alpha}(\p,\sigma)$ are given by (see eqs.~(\ref{eq:aat}) and (\ref{eq:bat}) for more details)
\begin{eqnarray}
&& a_{\alpha}(\p,\sigma,t)=\sum_{\beta,\sigma'}\left[U_{\alpha\beta}(\p,\sigma,t)
a_{\beta}(\p,\sigma)+V_{\alpha\beta}(\p,\sigma,\sigma',t)
b^{\dag}_{\beta}(-\p,\sigma')\right], \label{eq:aatt}\\
&& b_{\alpha}(\p,\sigma,t)=\sum_{\beta,\sigma'}\left[U_{\alpha\beta}(\p,\sigma,t)
b_{\beta}(\p,\sigma)+V_{\alpha\beta}(\p,\sigma,\sigma',t) \label{eq:batt}
a^{\dag}_{\beta}(-\p,\sigma')\right].
\end{eqnarray}
The right-hand side of eqs.~(\ref{eq:aatt}) and (\ref{eq:batt}) are sums of operators for particles and anti-particles over spin-projections and flavors. It then follows that the field $\nu_{\alpha}(x)$ at time $t$ is also a sum of fields of different flavors (see eqs.~(\ref{eq:net} and (\ref{eq:nmt}))
\begin{eqnarray}
&& \nu_{e}(x)=\lambda_{e}(x)+\lambda_{\mu}(x), \label{eq:n1}\\
&& \nu_{\mu}(x)=\varrho_{e}(x)+\varrho_{\mu}(x). \label{eq:n2}
\end{eqnarray}
What the above equations tell us is that when oscillation is taken into account, flavor is not conserved~\cite{Blasone:2006jx}. %Equation~(\ref{eq:n1}) becomes important when 
This fact becomes important in the next section when we compute the neutron $\beta$ decay spectrum.

%------------------------------------------------------------------------------------------------------------
\section{Neutron beta decay}\label{exp}

In the Minkowski space-time at zero temperature, the $S$-matrix for particles of definite mass is well-defined and it is used to compute observables such as cross-sections and decay rates. However, it is not immediately obvious whether $S$-matrix is applicable to flavor eigenstates due to the effect of oscillations. To understand this issue more concretely, we recall that the $S$-matrix for the transition of multi-particle state $|A\rangle$ to $|B\rangle$ is defined as
\begin{equation}
S_{BA}= {_{\tiny{\mbox{out}}}\langle} B|A\rangle_{\tiny{\mbox{in}}}
\end{equation}
where $|A\rangle_{\tiny{\mbox{in}}}$ and $|B\rangle_{\tiny{\mbox{out}}}$ are the in- and out-states. They are related to the free states $|A\rangle_{0}$ and $|B\rangle_{0}$ in the limit far before and after the interactions
\begin{eqnarray}
&& \lim_{t\rightarrow-\infty}e^{-iHt}|A\rangle_{\mbox{\tiny{in}}}=\lim_{t\rightarrow-\infty}e^{-iH_{0}t}|A\rangle_{0},\label{eq:ina}\\
&&\lim_{t\rightarrow +\infty}e^{-iHt}|B\rangle_{\mbox{\tiny{out}}}=\lim_{t\rightarrow +\infty}e^{-iH_{0}t}|B\rangle_{0}\label{eq:outa}
\end{eqnarray}
where $H=H_{0}+V(t)$ is the full Hamiltonian and it reduces to $H_{0}$ in the limit $t\rightarrow\pm\infty$. From eqs.~(\ref{eq:ina}) and (\ref{eq:outa}), the $S$-matrix can then be written as
\begin{equation}
S_{BA}= {_{0}\langle B}|S|A\rangle_{0}
\end{equation}
with
\begin{equation}
S=\lim_{t\rightarrow-\infty} \left[(e^{-iH_{0}t}e^{iHt}) (e^{iHt}e^{-iH_{0}t})\right]
\end{equation}
 which can be expanded using the Dyson series
\begin{eqnarray}
S&=&1-i\int_{-\infty}^{+\infty}dt\, V(t)+
(-i)^{2}\int_{-\infty}^{+\infty}dt_{1}\int_{-\infty}^{t_{1}}dt_{2}V(t_{1})V(t_{2})+\cdots.
%\nonumber \\
%&&+(-i)^{3}\int_{-T/2}^{T/2}dt_{1}\int_{-T/2}^{t_{1}}dt_{2}\int_{-T/2}^{t_{2}}dt_{3}V(t_{1})V(t_{2})V(t_{3})+\cdots.
\end{eqnarray}
For mass eigenstates such as the electrons, eqs.~(\ref{eq:ina}) and (\ref{eq:outa}) are well-defined since the identity of the electron remains the same throughout the time evolution. More specifically, for electrons, the associated mass and charge operators commute with the Hamiltonian and are hence conserved.

The flavor eigenstates defined in eqs.~(\ref{eq:fs1}) and (\ref{eq:fs2}) do not satisfy eqs.~(\ref{eq:ina}) and (\ref{eq:outa}) due to the effects of oscillation. That is, the flavor operator does not commute with the Hamiltonian so it is not conserved. 
%To overcome this difficulty, instead of taking $t\rightarrow\pm\infty$, one may take $t\rightarrow\pm T$ where $T\ll 1/(E_{1}+E_{2})$ so the oscillation is negligible but large enough such that $V(|T|)\rightarrow 0$. But this is not a satisfactory solution. At finite $T$, it is not possible to impose energy conservation. Without it, it is not clear how physical observables can be extracted from the theory.
This problem does not arise in the SM or its extension with massive neutrinos where they interact as mass eigenstates. The mass eigenstates do not oscillate so they are compatible with the $S$-matrix. The actual flavor oscillation probabilities can be derived by taking into account the associated flavor of the leptons produced during production and detection~\cite{Giunti:1991cb,Giunti:1993se,Giunti:2007ry}. As for the BV theory, some preliminary results on the interactions of flavor eigenstates can be found in refs.~\cite{Li:2006qt,Blasone:2006jx} but the calculations have not been completed. 

In this section, we compute the neutron $\beta$-decay spectrum using the BV theory with two neutrinos. The $S$-matrix is computed using the standard prescription. The out-going flavor eigenstates are free and the flavor fields are in the interacting picture. The only complication arises from the time-evolution of the flavor operators given in eqs.~(\ref{eq:aatt}) and (\ref{eq:batt}). The supposed ambiguity in the $T\rightarrow\infty$ limit for the flavor eigenstates does not pose any technical difficulties in performing the computations.  We show that in the limit $T\rightarrow\infty$, the spectrum naturally incorporates the effect of flavor oscillations. For relativistic neutrinos, the results agree with the SM.

To compute the neutron $\beta$ decay spectrum, we use the following effective Lagrangian
\begin{equation}
\mathscr{L}_{\mbox{\tiny{eff}}}=\frac{G_{F}}{\sqrt{2}}\left[\,\overline{\psi}_{e}\gamma^{\mu}(I-\gamma^{5})\nu_{e}\right]\left[
V_{ud}\overline{\psi}_{u}\gamma_{\mu}(f-\gamma^{5}g)\psi_{d}\right]
\end{equation}
where $\psi_{e}(x)$, $\psi_{u}(x)$ and $\psi_{d}(x)$ are the electron, up and down quark fields respectively while $f$ and $g$ are the form factors. As we have discussed in the previous section, since $\nu_{e}(x)$ is a sum of $\lambda_{e}(x)$ and $\lambda_{\mu}(x)$, there are two possible decay channels, $n\rightarrow p^{+}+e^{-}+\overline{\nu}_{\alpha}$ where $\alpha=e,\mu$. To the leading order, the $S$-matrix is given by
\begin{eqnarray}
S_{\alpha}&=&-\frac{iG_{F}}{\sqrt{2}\mathcal{N}}\left[(2\pi)^{3}\delta^{3}(\p_{e}+\p_{\nu}+\p_{p}-\p_{n})\right]\int^{T/2}_{-T/2}dt\,e^{i(E_{e}+E_{p}-E_{n})t}\nonumber\\
&&\times\sum_{\sigma}
\left[\overline{u}(\p_{e},\sigma_{e})\gamma^{\mu}(I-\gamma^{5})\mathcal{V}_{\alpha}(\p_{\nu},\sigma,\sigma_{\nu},t)\right]
\left[V_{ud}\,\overline{u}(\p_{p},\sigma_{p})\gamma^{\mu}(f-\gamma^{5}g)
u(\p_{n},\sigma_{n})\right]
\end{eqnarray}
where $\mathcal{V}_{\alpha}(\p_{\nu},\sigma,\sigma_{\nu},t)$ is given by eqs.~(\ref{eq:ve}) or (\ref{eq:vm}) and $\mathcal{N}$ is an energy-dependent normalization factor
\begin{equation}
\mathcal{N}=\frac{1}{(2\pi)^{6}(16E_{n}E_{p}E_{e}E_{1})^{1/2}}.
\end{equation}
% from Here, it is helpful to remind ourselves that $E_{i}=\sqrt{|\p_{\nu}|^{2}+m^{2}_{i}}$ where $m_{i}$ is one of the neutrino mass eigenstates. 

%------------------------------------------------------------------------------------------------------------------------------
\subsection{$n\rightarrow p^{+}+e^{-}+\overline{\nu}_{e}$}
%------------------------------------------------------------------------------------------------------------------------------

We start with the process $n\rightarrow p^{+}+e^{-}+\overline{\nu}_{e}$. Substituting eq.~(\ref{eq:ve}) into $S_{e}$, the $S$-matrix reads
%\footnote{From now onwards in the equations, we will neglect the 
%$\mathbf{p}_{\nu}$ and $\sigma_{\nu}$ arguments for $U^{2}$ and $V^{2}$.}
\begin{eqnarray}
S_{e}&=&-\frac{iG_{F}}{\sqrt{2}\mathcal{N}}\left[(2\pi)^{3}\delta^{3}(\p_{e}+\p_{\nu}+\p_{p}-\p_{n})\right](2\pi)\overline{u}(\p_{e},\sigma_{e})\gamma^{\mu}(I-\gamma^{5})\nonumber\\
&&\Big{\lbrace}\cos^{2}\theta v_{1}(\p_{\nu},\sigma_{\nu})\delta_{T}(E_{p}+E_{e}+E_{1}-E_{n})\nonumber\\
&&+\sin^{2}\theta\left[U^{2} v_{1}(\p_{\nu},\sigma_{\nu})+ U\sum_{\sigma}V(-\p_{\nu},\sigma,\sigma_{\nu})u_{1}(-\p_{\nu},\sigma)\right]
\delta_{T}(E_{p}+E_{e}+E_{2}-E_{n})\nonumber\\
&&+\sin^{2}\theta\left[V^{2}v_{1}(\p_{\nu},\sigma_{\nu})-U
\sum_{\sigma}V(-\p_{\nu},\sigma,\sigma_{\nu})u_{1}(-\p_{\nu},\sigma)\right]
\delta_{T}(E_{p}+E_{e}-E_{2}-E_{n})\Big{\rbrace}\nonumber\\
&&\times V_{ud}\overline{u}(\p_{p},\sigma_{p})\gamma_{\mu}(f-\gamma^{5}g)
u(\p_{n},\sigma_{n})
\end{eqnarray}
where 
\begin{equation}
\delta_{T}(E-E')=\frac{1}{2\pi}\int^{T/2}_{-T/2}dt \exp\left[i(E-E')t\right]
\end{equation}
Here, we may proceed with the calculation by taking the limit $T\rightarrow\infty$ but we find it more instructive to rewrite the $S$-matrix as
\begin{eqnarray}
S_{e}&=&-\frac{iG_{F}}{\sqrt{2}\mathcal{N}}\left[(2\pi)^{3}\delta^{3}(\p_{e}+\p_{\nu}+\p_{p}-\p_{n})\right]\overline{u}(\p_{e},\sigma_{e})\gamma^{\mu}(I-\gamma^{5})\nonumber\\
&&\times\left[F(T)v_{1}(\p_{\nu},\sigma_{\nu})+UG(T)\sum_{\sigma}
V(-\p_{\nu},\sigma,\sigma_{\nu})u_{1}(-\p_{\nu},\sigma)\right]\nonumber\\
&&\times V_{ud}\overline{u}(\p_{p},\sigma_{p})\gamma_{\mu}(f-\gamma^{5}g)
u(\p_{n},\sigma_{n}) \label{eq:se}
\end{eqnarray}
and take the limit later. The functions $F(T)$ and $G(T)$ are defined as
\begin{eqnarray}
\hspace{-0.5cm}F(T)&\equiv& +2\pi\cos^{2}\theta \delta_{T}(E_{p}+E_{e}+E_{1}-E_{n})\nonumber\\
&&+2\pi\sin^{2}\theta\left[U^{2}\delta_{T}(E_{p}+E_{e}+E_{2}-E_{n})
+V^{2}\delta_{T}(E_{p}+E_{e}-E_{2}-E_{n})\right],
\end{eqnarray}
\begin{eqnarray}
G(T)&\equiv&2\pi\sin^{2}\theta\left[ \delta_{T}(E_{p}+E_{e}+E_{2}-E_{n})
-\delta_{T}(E_{p}+E_{e}-E_{2}-E_{n})\right].
\end{eqnarray}
For small $T$, we find that $F(T)\approx 1$ and $G(T)\approx 0$ so the $S$-matrix takes a similar from the SM. To compute the decay rate, it is convenient to write the $S$-matrix as
\begin{eqnarray}
S_{e}&=&-2\pi i\delta^{3}(\p_{p}+\p_{e}+\p_{\nu}-\p_{n})K^{\mu}M_{\mu}
\end{eqnarray}
where 
\begin{eqnarray}
&&M_{\mu}=\frac{(2\pi)^{2}G_{F}}{\sqrt{2}\mathcal{N}}V_{ud}
\overline{u}(\p_{p},\sigma_{p})\gamma_{\mu}(f-\gamma^{5}g)u(\p_{n},\sigma_{n}),\\
&&K^{\mu}=\overline{u}(\p_{e},\sigma_{e})\gamma^{\mu}(I-\gamma^{5})
\left[Fv_{1}(\p_{\nu},\sigma_{\nu})+UG\sum_{\sigma}V(-\p_{\nu},\sigma,\sigma_{\nu})u_{1}(-\p_{\nu},\sigma)\right].
\end{eqnarray}
Summing over the spin degrees of freedom, the averaged differential decay rate is given by (see app.~\ref{tr})
\begin{eqnarray}
d\Gamma_{\tiny{\mbox{avg}}}(\overline{\nu}_{e})&=&\frac{1}{2}\frac{(2\pi)^{2}}{T}\sum_{\tiny{\mbox{spins}}}|K^{\mu}M_{\mu}|^{2}\delta^{3}(\p_{p}+\p_{e}+\p_{\nu}-\p_{n})d^{3}p_{p}d^{3}p_{e}d^{3}p_{\nu}.
\end{eqnarray}

Here, we take the neutron to be at rest. Since $m_{n}\approx m_{p}$ and $m_{p}\gg m_{i},m_{e}$, we may take the approximation $|\p_{p}|\approx0$ ignoring the proton recoil. The spin-sums given in eqs.~(\ref{eq:k1})-(\ref{eq:k3}) simplify and their sum is
\begin{eqnarray}
\sum_{\tiny{\mbox{spins}}}|K^{\mu}M_{\mu}|^{2}&=&
\frac{2G^{2}_{F}V^{2}_{ud}}{(2\pi)^{8}(E_{e}E_{1})}
\Big{\{}(f^{2}+3g^{2})\left[F^{2}+2U^{2}V^{2}G^{2}\right]
E_{e}E_{1}\nonumber\\
&&+(f^{2}-g^{2})\left[F^{2}-2U^{2}V^{2}G^{2}\right]
m_{n}m_{p}(\p_{e}\cdot\p_{\nu})\Big{\}} \nonumber\\
&&+\frac{2G^{2}_{F}V^{2}_{ud}}{(2\pi)^{8}(E_{e}E_{1})}
\frac{m_{1}\left[|\p_{\nu}|^{2}+(m_{1}+E_{1})(m_{2}+E_{2})\right](m_{1}-m_{2}+E_{1}-E_{2})}{2E_{1}E_{2}(m_{1}+E_{1})(m_{2}+E_{2})}\nonumber\\
&&\times (f^{2}-g^{2})FG(\p_{e}\cdot\p_{\nu}).
\end{eqnarray}
Using the approximation $|\p_{p}|\approx 0$, the average differential decay rate becomes
\begin{equation}
d\Gamma_{\tiny{\mbox{avg}}}(\overline{\nu}_{e})=\frac{1}{2}\frac{(2\pi)^{2}}{T}
\sum_{\tiny{\mbox{spins}}}|K^{\mu}M_{\mu}|^{2} d^{3}p_{e}d^{3}p_{\nu}.
\end{equation}
Setting up a spherical coordinate system with respect to $\p_{\nu}$, the terms containing $\p_{e}\cdot\p_{\nu}$ vanish after integration leaving us with
\begin{equation}
d\Gamma_{\tiny{\mbox{avg}}}(\overline{\nu}_{e})=\frac{G^{2}_{F}V^{2}_{ud}}{4\pi^{4}T}
(f^{2}+3g^{2})|\p_{e}|^{2}|\p_{\nu}|^{2}\left[F^{2}+2U^{2}V^{2}
G^{2}\right]dp_{e}dp_{\nu}.
\end{equation}
In order to proceed further, we need to take care of the $\delta_{T}$ functions. At finite time, $F^{2}(T)$ and $G^{2}(T)$ consists of various products of $\delta_{T}$ functions. In the limit $T\rightarrow\infty$, only the diagonal terms remain and $\delta_{T}$ becomes the Dirac $\delta$ function
\begin{eqnarray}
\lim_{T\rightarrow\infty}F^{2}(T)&=&\lim_{T\rightarrow\infty}(2\pi T)\Big{\{}\cos^{4}\theta\,\delta(E_{e}+E_{1}+m_{p}-m_{n})\nonumber\\
&&+\sin^{4}\theta\left[U^{4}\delta(E_{e}+E_{2}+m_{p}-m_{n})+V^{4}\delta(E_{e}-E_{2}+m_{p}-m_{n})\right]\Big{\}},
\end{eqnarray}
\begin{eqnarray}
 \lim_{T\rightarrow\infty}G^{2}(T)=\lim_{T\rightarrow\infty}(2\pi T)\sin^{4}\theta\left[\delta(E_{e}+E_{2}+m_{p}-m_{n})+\delta(E_{e}-E_{2}+m_{p}-m_{n})\right].
\end{eqnarray}
Therefore, the spectrum for $n\rightarrow p^{+}+e^{-}+\overline{\nu}_{e}$ is
\begin{eqnarray}
\frac{d\Gamma_{\tiny{\mbox{avg}}}}{dE_{e}}(\overline{\nu}_{e})&=&\frac{G^{2}_{F}V^{2}_{ud}}{2\pi^{3}}
(f^{2}+3g^{2})|\p_{\nu}||\p_{e}|E_{e}\Big{\{}\int_{E_{1}>m_{1}} dE_{1}\,E_{1}\cos^{4}\theta \delta(E_{e}+E_{1}+m_{p}-m_{n})\nonumber\\
&&+\int_{E_{2}>m_{2}}dE_{2}\,E_{2}\sin^{4}\theta \Big{[}\left(U^{4}+2U^{2}V^{2}\right)
\delta(E_{e}+E_{1}+m_{p}-m_{n})\nonumber\\
&&\hspace{1.6cm}+\left(V^{4}+2U^{2}V^{2}\right)\delta(E_{e}-E_{2}+m_{p}-m_{n})\Big{]}
\Big{\}}.
\end{eqnarray}
The coefficients $U$ and $V$ are functions of $E_{1}$ and $E_{2}$. To perform the integration over energies, we rewrite them as functions of one another
\begin{equation}
E_{1}=\sqrt{E^{2}_{2}-\Delta m^{2}},\quad
E_{2}=\sqrt{E^{2}_{1}+\Delta m^{2}}.
\end{equation}
The  $\delta(E_{e}-E_{2}+m_{p}-m_{n})$  term in the spectrum does not contribute since $E_{i}>0$ for $m_{i}>0$ but the $\delta$-function forces $E_{2}$ to be $E_{e}-(m_{n}-m_{p})\leq 0$. Therefore, this term identically vanishes. Finally, integrating over $E_{1}$ and $E_{2}$, we obtain
\begin{eqnarray}
\frac{d\Gamma_{\tiny{\mbox{avg}}}}{dE_{e}}(\overline{\nu}_{e})&=&\frac{G^{2}_{F}V^{2}_{ud}}{2\pi^{3}}
(f^{2}+3g^{2})|\p_{e}|E_{e}(m_{n}-m_{p}-E_{e})\nonumber\\
&&\times\left[
\cos^{4}\theta\sqrt{(m_{n}-m_{p}-E_{e})^{2}-m^{2}_{1}}
+\sin^{4}\theta(U^{4}+2U^{2}V^{2})
\sqrt{(m_{n}-m_{p}-E_{e})^{2}-m^{2}_{2}}\right]\nonumber\\
\end{eqnarray}
with the understanding that
\begin{equation}
E_{1}\rightarrow\sqrt{(m_{n}-m_{p}-E_{e})^{2}-\Delta m^{2}},\quad
E_{2}\rightarrow (m_{n}-m_{p}-E_{e})
\end{equation}
for the coefficients $U$ and $V$.

%------------------------------------------------------------------------------------------------------------------------------
\subsection{$n\rightarrow p^{+}+e^{-}+\overline{\nu}_{\mu}$}
%------------------------------------------------------------------------------------------------------------------------------

At the leading order, the $S$-matrix for $n\rightarrow p^{+}+e^{-}+\overline{\nu}_{\mu}$ is
\begin{eqnarray}
S_{\mu}&=&-\frac{iG_{F}}{\sqrt{2}\mathcal{N}}\left[(2\pi)^{3}\delta^{3}(\p_{e}+\p_{\nu}+\p_{p}-\p_{n})\right](2\pi)\cos\theta\sin\theta\,
\overline{u}(\p_{e},\sigma_{e})\gamma^{\mu}(I-\gamma^{5})\nonumber\\
&&\times\Big{\lbrace}
\left[U v_{1}(\p_{\nu},\sigma_{\nu})+\sum_{\sigma}V(-\p_{\nu},\sigma,\sigma_{\nu})u_{1}(-\p_{\nu},\sigma)\right]
\delta_{T}(E_{p}+E_{e}+E_{2}-E_{n})\nonumber\\
&&-\sum_{\sigma}V(-\p_{\nu},\sigma,\sigma_{\nu})u_{1}(-\p_{\nu},\sigma)
\delta_{T}(E_{p}+E_{e}-E_{1}-E_{n})\nonumber\\
&&-Uv_{1}(\p_{\nu},\sigma_{\nu})
\delta_{T}(E_{p}+E_{e}+E_{1}-E_{n})\Big{\rbrace}.
\end{eqnarray}
We rewrite the $S$-matrix as
\begin{eqnarray}
S_{\mu}&=&-\frac{iG_{F}}{\sqrt{2}\mathcal{N}}\left[(2\pi)^{3}\delta^{3}(\p_{e}+\p_{\nu}+\p_{p}-\p_{n})\right]\cos\theta\sin\theta\,
\overline{u}(\p_{e},\sigma_{e})\gamma^{\mu}(I-\gamma^{5})\nonumber\\
&&\times\left[UJ(T)v_{1}(\p_{\nu},\sigma_{\nu})
+K(T)\sum_{\sigma}V(-\p_{\nu},\sigma,\sigma_{\nu})u_{1}(-\p_{\nu},\sigma)\right]\nonumber\\
&&\times V_{ud}\overline{u}(\p_{p},\sigma_{p})\gamma_{\mu}(f-\gamma^{5}g)u(\p_{n},\sigma_{n})\label{eq:smu}
\end{eqnarray}
where
\begin{eqnarray}
&& J(T)\equiv 2\pi\left[\delta_{T}(E_{p}+E_{e}+E_{2}-E_{n})
-\delta_{T}(E_{p}+E_{e}+E_{1}-E_{n})\right],\\
&& K(T)\equiv 2\pi\left[\delta_{T}(E_{p}+E_{e}+E_{2}-E_{n})
-\delta_{T}(E_{p}+E_{e}-E_{1}-E_{n})\right].
\end{eqnarray}
At short time, $J(T)\approx0$, $K(T)\approx0$ so the amplitude for the flavor-violating process vanishes. But here, we are interested in the long-term behavior where $T\rightarrow\infty$. Comparing eq.~(\ref{eq:se}) to (\ref{eq:smu}), we notice that $S_{e}\rightarrow S_{\mu}$ when $F(T)\rightarrow UJ(T)$ and $G(T)\rightarrow K(T)/U$. Therefore,
\begin{equation}
d\Gamma_{\tiny{\mbox{avg}}}(\overline{\nu}_{\mu})=\frac{G^{2}_{F}V^{2}_{ud}}{4\pi^{4}T}
(f^{2}+3g^{2})|\p_{e}|^{2}|\p_{\nu}|^{2}\left[U^{2}J^{2}+2V^{2}K\right]dp_{e}dp_{\nu}.
\end{equation}
Similar to $F(T)$ and $G(T)$, in the limit $T\rightarrow\infty$, we have
\begin{eqnarray}
&& J^{2}(T)=\lim_{T\rightarrow\infty}(2\pi T)\left[\delta(E_{p}+E_{e}+E_{2}-E_{n})
+\delta(E_{p}+E_{e}+E_{1}-E_{n})\right],\\
&& K^{2}(T)=\lim_{T\rightarrow\infty}(2\pi T)\left[\delta(E_{p}+E_{e}+E_{2}-E_{n})
-\delta(E_{p}+E_{e}-E_{1}-E_{n})\right].
\end{eqnarray}
Performing the integration over $|\p_{\nu}|$, the spectrum for $n\rightarrow p^{+}+e^{-}+\overline{\nu}_{\mu}$ is
\begin{equation}
\frac{d\Gamma_{\tiny{\mbox{avg}}}}{dE_{e}}(\overline{\nu}_{\mu})=
\frac{d\Gamma^{(1)}_{\tiny{\mbox{avg}}}}{dE_{e}}(\overline{\nu}_{\mu})+\frac{d\Gamma^{(2)}_{\tiny{\mbox{avg}}}}{dE_{e}}(\overline{\nu}_{\mu})
\end{equation}
where
\begin{eqnarray}
\frac{d\Gamma^{(1)}_{\tiny{\mbox{avg}}}}{dE_{e}}(\overline{\nu}_{\mu})&=&
\frac{G^{2}_{F}V^{2}_{ud}}{2\pi^{3}}(f^{2}+3g^{2})U^{2}|\p_{e}|E_{e}(\cos\theta\sin\theta)^{2}\nonumber\\
&&\times\left[(m_{n}-m_{p}-E_{e})\sqrt{(m_{n}-m_{p}-E_{e})^{2}-m^{2}_{1}}\right],
\end{eqnarray}
\begin{equation}
E_{1}\rightarrow(m_{n}-m_{p}-E_{e}),\quad
E_{2}\rightarrow\sqrt{(m_{n}-m_{p}-E_{e})^{2}+\Delta m^{2}}
\end{equation}
\begin{eqnarray}
\frac{d\Gamma^{(2)}_{\tiny{\mbox{avg}}}}{dE_{e}}(\overline{\nu}_{\mu})&=&
\frac{G^{2}_{F}V^{2}_{ud}}{2\pi^{3}}(f^{2}+3g^{2})(U^{2}+2V^{2})|\p_{e}|E_{e}(\cos\theta\sin\theta)^{2}\nonumber\\
&&\times\left[(m_{n}-m_{p}-E_{e})\sqrt{(m_{n}-m_{p}-E_{e})^{2}-m^{2}_{2}}\right],
\end{eqnarray}
\begin{equation}
E_{1}\rightarrow\sqrt{(m_{n}-m_{p}-E_{e})^{2}-\Delta m^{2}},\quad
E_{2}\rightarrow(m_{n}-m_{p}-E_{e}).
\end{equation}
The substitutions of $E_{1}$ and $E_{2}$ are to be made for $U$ and $V$.

\begin{figure}
\resizebox{1\hsize}{!}{\includegraphics*{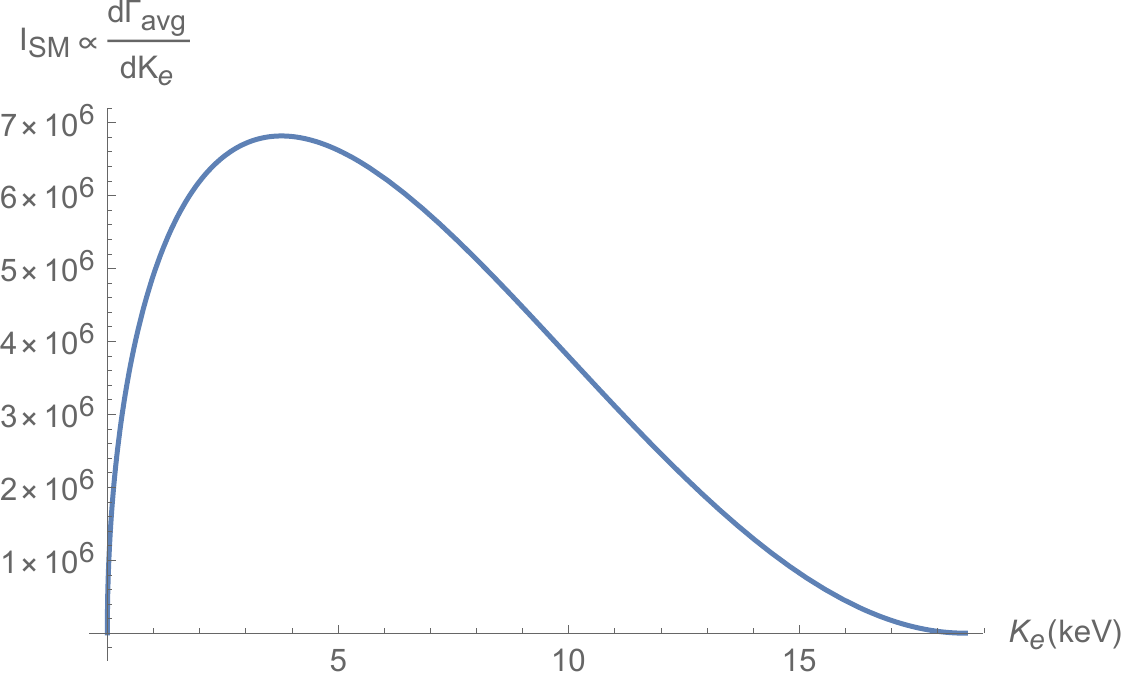}}
\caption{The intensity of Tritium $\beta$ decay for
$\ce{^{3}_{1}H}\rightarrow \ce{^{3}_{2}He}+e^{-}+\bar{\nu}_{e}$ in the SM with massless neutrino. The $x$-axis is the electron kinetic energy $K_{e}=E_{e}-m_{e}$. The $y$-axis is the intensity which is proportional to $d\Gamma_{\tiny{\mbox{avg}}}/dK_{e}$. The proportionality constant that we have factored out is $[G^{2}_{F}V^{2}_{ud}/(2\pi)^{3}]f'$ where $f'$ is the appropriate nuclear matrix element.}\label{fig:ism}
\end{figure}
\begin{figure}
\resizebox{1\hsize}{!}{\includegraphics{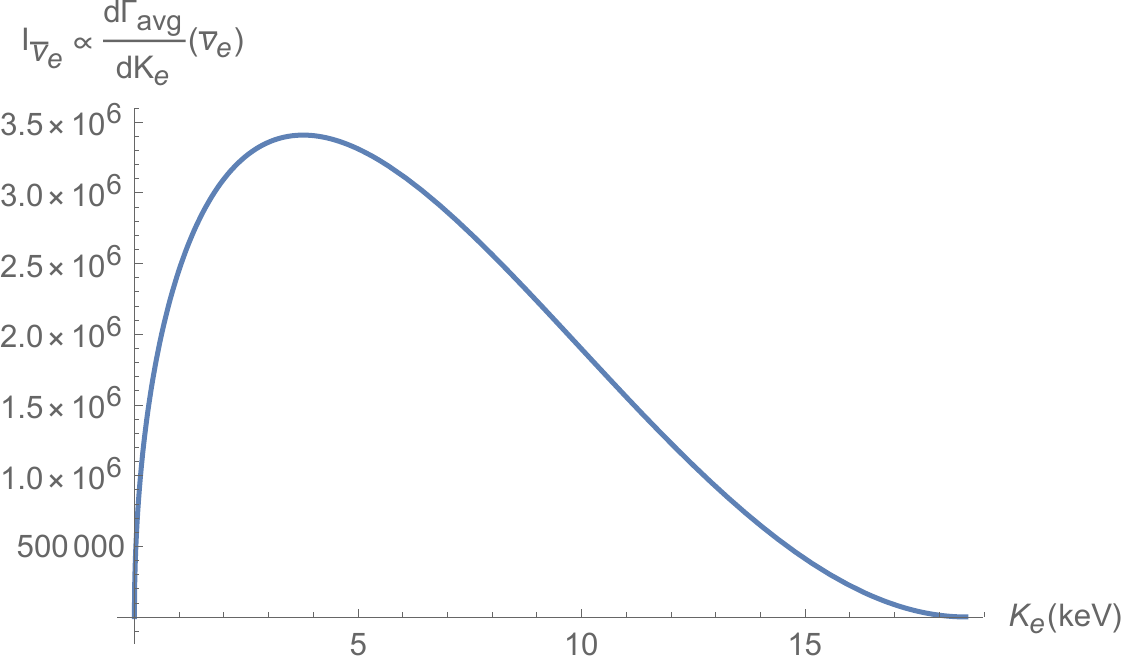}}
\label{fig:subim2}
\resizebox{1\hsize}{!}{\includegraphics{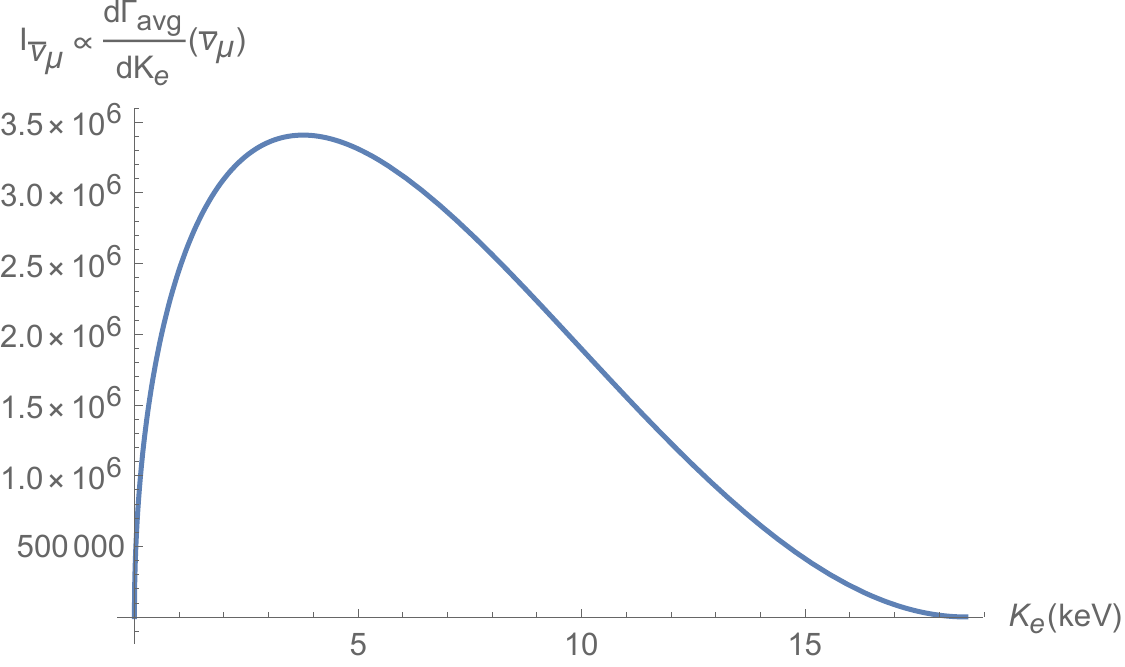}}
\label{fig:subim3}
\caption{The intensity of Tritium $\beta$ decay for
$\ce{^{3}_{1}H}\rightarrow \ce{^{3}_{2}He}+e^{-}+\bar{\nu}_{\alpha}$ where $\alpha=e,\mu$ with flavor eigenstates. We chose $m_{1}=0.1\,\mbox{eV}$, $m_{2}=0.2\,\mbox{eV}$ and $\sin^{2}2\theta=\frac{1}{2}$. The proportionality constant that we have factored out is the same as fig.~\ref{fig:ism}}
\label{fig:image2}
\end{figure}

The neutron $\beta$-decay spectrum in the BV theory with two neutrinos is given by
\begin{equation}
\frac{d\Gamma_{\tiny{\mbox{avg}}}}{dE_{e}}=
\frac{d\Gamma_{\tiny{\mbox{avg}}}}{dE_{e}}(\overline{\nu}_{e})+\frac{d\Gamma_{\tiny{\mbox{avg}}}}{dE_{e}}(\overline{\nu}_{\mu}).
\end{equation}
The important feature is that the flavor-violating process contributes to the decay process. if we ignore the mixing angles, the contributions from the flavor-conserving and violating process are of the same order. This result is a direct consequence of the fact that the BV theory incorporates the effect of flavor oscillations. When we perform the time integration and take the limit $T\rightarrow\infty$, we are effectively averaging the oscillation probability. To see this, we note that the effects of non-relativistic neutrinos are only important near the tail end of the spectrum where $E_{e}\approx m_{n}-m_{p}$ so we may take neutrinos to be relativistic where $|\p_{\nu}|\gg m_{i}$. In this limit $U\approx 1$ and $V\approx 0$ so we obtain
\begin{eqnarray}
&&\frac{d\Gamma_{\tiny{\mbox{avg}}}}{dE_{e}}(\overline{\nu}_{e})=
\left(1-\frac{1}{2}\sin^{2}2\theta\right)\frac{G^{2}_{F}V^{2}_{ud}}{2\pi^{3}}(f^{2}+3g^{2})
|\p_{e}|E_{e}(m_{n}-m_{p}-E_{e})^{2}+\cdots,\\
&&\frac{d\Gamma_{\tiny{\mbox{avg}}}}{dE_{e}}(\overline{\nu}_{\mu})=
\left(\frac{1}{2}\sin^{2}2\theta\right)\frac{G^{2}_{F}V^{2}_{ud}}{2\pi^{3}}(f^{2}+3g^{2})
|\p_{e}|E_{e}(m_{n}-m_{p}-E_{e})^{2}+\cdots.
\end{eqnarray}
These two spectra are simply the SM spectrum for massless neutrino multiplied by the average oscillation probability
\begin{equation}
\frac{d\Gamma_{\tiny{\mbox{avg}}}}{dE_{e}}(\overline{\nu}_{\alpha})=
\mathcal{Q}^{\tiny{\mbox{avg}}}_{e\rightarrow\alpha}
\frac{d\Gamma^{(m_{\nu}=0)}_{\tiny{\mbox{SM,avg}}}}{dE_{e}}(\overline{\nu}_{e})+\cdots
\label{eq:bvp}
\end{equation}
where
\begin{eqnarray}
&&\mathcal{Q}^{\tiny{\mbox{avg}}}_{e\rightarrow e}=1-\frac{1}{2}\sin^{2}2\theta,\label{eq:avge} \\
&&\mathcal{Q}^{\tiny{\mbox{avg}}}_{e\rightarrow\mu}=\frac{1}{2}\sin^{2}2\theta \label{eq:avgm}
\end{eqnarray}
and
\begin{equation}
\frac{d\Gamma^{(m_{\nu}=0)}_{\tiny{\mbox{SM,avg}}}}{dE_{e}}(\overline{\nu}_{e})=
\frac{G^{2}_{F}V^{2}_{ud}}{2\pi^{3}}(f^{2}+3g^{2})|\p_{e}|E_{e}(m_{n}-m_{p}-E_{e})^{2}
\end{equation}
thus verifying our claim. Summing over the flavors, we obtain
\begin{equation}
\frac{d\Gamma_{\tiny{\mbox{avg}}}}{dE_{e}}=\sum_{\alpha}\frac{d\Gamma_{\tiny{\mbox{avg}}}}{dE_{e}}(\overline{\nu}_{\alpha})=
\frac{d\Gamma^{(m_{\nu}=0)}_{\tiny{\mbox{SM,avg}}}}{dE_{e}}(\overline{\nu}_{e})+\cdots.
\end{equation}

The spectrum that we have computed in the $T\rightarrow\infty$ limit deserves careful elaboration. In $S$-matrix calculations, this limit is connected to the condition that far before and after the interaction, the incoming and outgoing particles are free of interactions. Physically, this corresponds to placing the detector at some finite distance away from the source such that it is not affected by the interaction.  But since the BV theory naturally incorporates the effect of oscillations, the limit $T\rightarrow\infty$ contains another important piece of information that is encoded in  eq.~(\ref{eq:bvp}). Specifically, for relativistic neutrinos, the factorization of the spectra suggests that as $T\rightarrow\infty$, the oscillation probabilities tend to its average values. This result corresponds to the situation where the detector is placed at a distance from the source greater than the coherent length of oscillation $L_{\tiny{\mbox{coh}}}$~\cite{Giunti:1997wq,Giunti:2002xg,Blasone:2002wp}. In the case of two neutrinos, the oscillation probabilities tend to eqs.~(\ref{eq:avge}) and (\ref{eq:avgm}). Therefore, to test the BV theory based on the spectra, the detector must be placed at a distance greater than $L_{\tiny{\mbox{coh}}}$ from the source. In practice, this is very challenging since to the best of our knowledge, present experiments cannot distinguish the suppression of oscillation probabilities due to $L_{\tiny{\mbox{coh}}}$ from the process of averaging as a result of uncertainties in distance and energy measurements. Additionally, measuring the electron kinetic energy to a precision of the order $0.1\,\mbox{eV}$ is also very challenging. %Therefore, additional efforts from the experimental side are required to control the neutrino production and detection process. 
Nevertheless, it is encouraging that at the leading order, the neutron $\beta$ decay spectrum obtained using the BV theory is in agreement with the SM.  

\begin{figure}
\resizebox{1\hsize}{!}{\includegraphics*{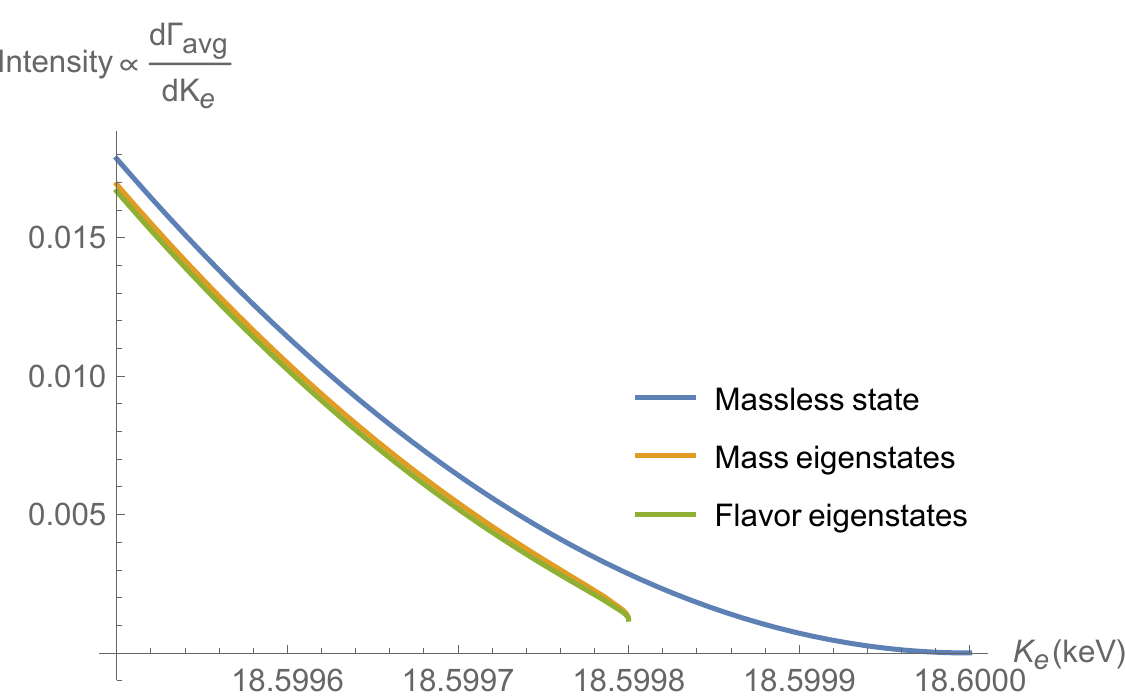}}
\caption{The intensity of Tritium $\beta$ decay near its end point of $Q_{T}\approx 18.6\,\mbox{keV}$ where the neutrinos interact as massless, mass and flavor eigenstates.  For the latter (the yellow and green lines), we chose $m_{1}=0.1\,\mbox{eV}$, $m_{2}=0.2\,\mbox{eV}$ and $\sin^{2}2\theta=\frac{1}{2}$. The proportionality constant that we have factored out is the same as fig.~\ref{fig:ism}.}\label{spec}
\end{figure}
\begin{figure}
\resizebox{1\hsize}{!}{\includegraphics*{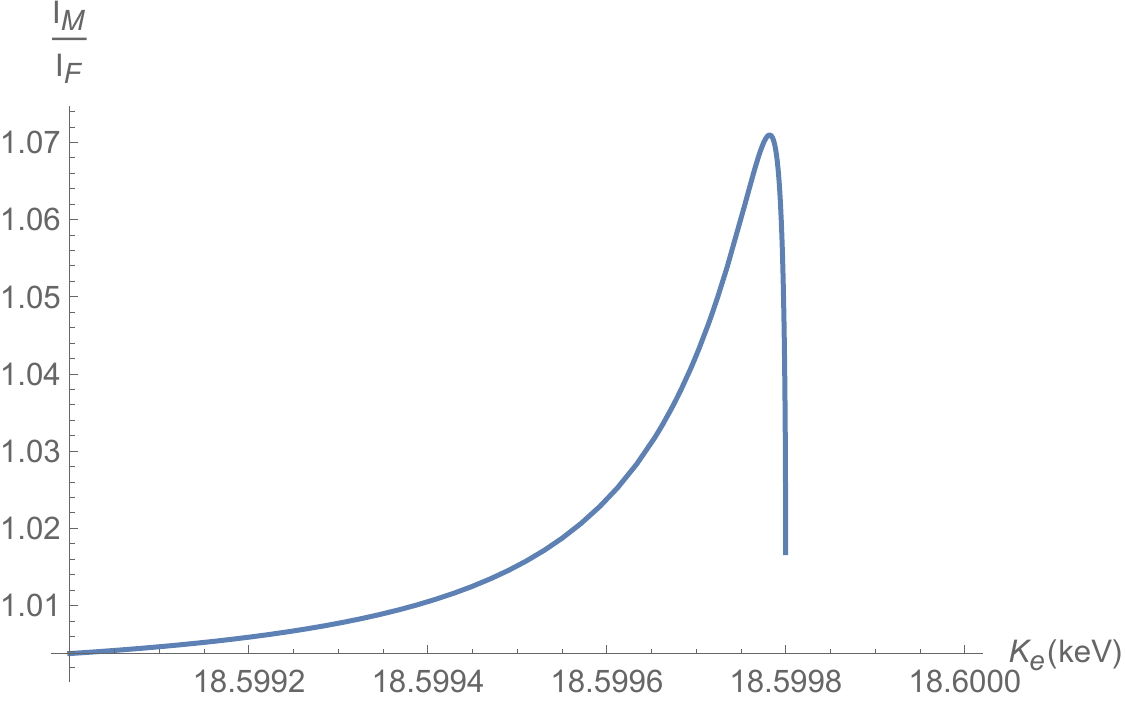}}
\caption{In this plot, $I_{M}=I_{1}+I_{2}$ and $I_{F}=I_{\overline{\nu}_{e}}+I_{\overline{\nu}_{\mu}}$ are the intensities for neutrinos as mass and flavor eigenstates respectively in Tritium $\beta$ decay. We chose $m_{1}=0.1\,\mbox{eV}$, $m_{2}=0.2\,\mbox{eV}$ and $\sin^{2}2\theta=\frac{1}{2}$.}\label{ratio}
\end{figure}
Although it may not be practical to perform the actual experiment, it is still instructive to compare predictions from different theories. In the case where the two neutrinos interact as mass eigenstates, the neutron decay channels are $n\rightarrow p^{+}+e^{-}+\overline{\nu}_{i}$ where $i=1,2$. The resulting spectrum is given by
\begin{eqnarray}
\frac{d\Gamma^{(m_{i})}_{\tiny{\mbox{avg}}}}{dE_{e}}&=&
\frac{G^{2}_{F}V^{2}_{ud}}{2\pi^{3}}(f^{2}+3g^{2})|\p_{e}|E_{e}(m_{n}-m_{p}-E_{e})\nonumber\\
&&\times\Big{[}\cos^{2}\theta\sqrt{(m_{n}-m_{p}-E_{e})^{2}-m^{2}_{1}}+\sin^{2}\theta\sqrt{(m_{n}-m_{p}-E_{e})^{2}-m^{2}_{2}}\Big{]}.
\end{eqnarray}

To see the discrepancies between the theories, it is more instructive to examine the spectrum of Tritium $\beta$ decay since its end point energy is $Q_{T}\approx 18.6\,\mbox{keV}$ as compared to $Q_{n}\approx 0.78\,\mbox{MeV}$ for neutrons. Having obtained the neutron $\beta$ decay spectrum, it is straightforward to obtain its counterpart for Tritium. One simply makes the substitutions $m_{n}\rightarrow m_{T}$, $m_{p}\rightarrow m_{He}$ and $f^{2}+3g^{2}$ with the appropriate nuclear matrix elements~\cite{Schiavilla:1998je,Long:2014zva}. In fig.~\ref{fig:image2}, due to the effect of oscillations, the intensities for each flavor are effectively identical. When compared with fig.~\ref{fig:ism}, we find $I_{\overline{\nu}_{\alpha},\tiny{\mbox{max}}}\approx \frac{1}{2}I_{\tiny{\mbox{SM, max}}}$. Nevertheless, the total intensities are in agreement $I_{\overline{\nu}_{e}}+I_{\overline{\nu}_{\mu}}\approx I_{\tiny{\mbox{SM}}}$ except towards the end point as evident from fig.~\ref{spec}.  The spectra for mass and flavor eigenstates have sharp cut off at 
$
K_{e}=Q_{T}-m_{2}\approx18.5998\,\mbox{keV}
$
 since beyond this value, the spectra become complex and physically meaningless. In fig.~\ref{ratio}, we have plotted the ratio of the intensity for neutrinos as mass and flavor eigenstates in Tritium $\beta$ decay. Near the end point, they differ by as much as 7\%.
 
 Strictly speaking, we should have performed the calculations with three neutrinos. Nevertheless, at the leading order, it seems reasonable to  expect eq.~(\ref{eq:bvp}) to hold for three or more neutrinos. The calculation for models with three or more neutrinos in the BV theory are considerably more complicated. For now, we leave this task and the proof of eq.~(\ref{eq:bvp}) for future investigations.

%--------------------------------------
\section{Conclusions}

In this paper, we computed the neutron $\beta$ decay spectrum for the BV theory with two neutrinos. Similar calculation was considered in refs.~\cite{Li:2006qt,Blasone:2006jx} but was incomplete. The main concern was whether flavor eigenstates are well-defined in the limit far before and after the interactions. The main results of this paper show that these limits are well-defined. For relativistic neutrinos, the neutron $\beta$ decay spectrum is in agreement with the SM. The flavor-violating process $n\rightarrow p^{+}+e^{-}+\overline{\nu}_{\mu}$ which contribute to the decay is due to the effect of oscillations as evident from eq.~(\ref{eq:bvp}). The factorization of the spectrum as given in eq.~(\ref{eq:bvp}) tells us that the $T\rightarrow\infty$ limit corresponds to an experimental set-up where the detector is placed at a distance greater than the coherent length of oscillation $L_{\tiny{\mbox{coh}}}$ from the source. Our results suggest that for the BV theory, it may not be practical to measure observables computed  via the standard $S$-matrix. Taking inspiration from ref.~\cite{Giunti:1993se}, a promising direction for future investigation is to see whether it is possible to obtain observables that are measured at finite distance when the production and detection processes are taken into account.

%Inspired by the KATRIN and PTOLEMY experiments, we proposed an experiment to study flavor oscillations at low-energy. However, this is only feasible for neutrinos with maximum energy less than a few keV and cannot  be used to study neutrinos in the sub-eV region.  To test the BV theory, another possibility is to study low-energy interactions where the neutrino momentum are around $|\p_{ij}|=\sqrt{m_{i}m_{j}}$. This is in fact possible for KATRIN and PTOLEMY. Therefore, an important task is to compute the neutrino capture and $\beta$-decay rates of Tritium with the flavor eigenstates. A realistic description requires us to take into account all three neutrinos. These calculations will be carried out in the future.

Assuming that future experiments are capable of measuring the effects of decoherence in neutrino oscillations, discrepancies between the BV theory and the SM in the neutron and Tritium $\beta$ decays can then be deciphered near the end point energy of the parent nucleus. For neutrons, this is approximately $Q_{n}\approx 1\,\mbox{MeV}$ which is about $10^{7}$ times larger than the neutrino mass so any effects are likely to be beyond the range of experimental sensitivities. A more suitable process that is currently under investigation by KATRIN is Tritium $\beta$ decay since it has a lower end point energy of $Q_{T}\approx 18.6\,\mbox{keV}$ and a simple nuclear structure. Accurate measurements of the electron kinetic energy near the Tritium end point would provide important information on the neutrino masses. From fig.~\ref{spec}, we see that for mass and flavor eigenstates, the cut-off provides information on the values of $m_{2}$. We expect this property to persist for models with three or more neutrinos. This means that an accurate measurement of the maximum kinetic energy of the electron will directly determine the maximum neutrino mass. The other masses can then be obtained through the mass squared difference.
%As evident from fig.~\ref{ratio}, the intensity between mass and flavour eigenstates can differ by as much as 7~\%.

Another important process to study in the future for the BV theory is neutrino capture by Tritium. This process is used by the PTOLEMY collaboration to detect the cosmic neutrino background. 
The capture of relic neutrinos by Tritium presents a unique opportunity to test the BV theory. Because the relic neutrino decouple from the cosmic plasma at approximately one second after the Big Bang, the distance travelled between time of decoupling and the present epoch would have exceeded coherent length of oscillation. Therefore, in the present epoch the relic neutrinos satisfy the condition required by the BV theory. The calculation presented in this paper includes only two neutrinos. A more realistic calculation with three neutrinos will be carried out in the future. 

Finally, the question we would like to answer is: How do neutrinos interact and how are they produced? Theories proposed by Schrock and BV present two different pictures but are of the same origin, namely the equation $\nu_{\alpha}(x)=\sum_{i}U_{\alpha i}\nu_{i}(x)$. The difference comes down to a choice between mass and flavor eigenstate basis when expanding $\nu_{\alpha}(x)$. Expansions in the mass eigenstate basis is conceptually and mathematically simpler to work with. But simplicity itself is not a valid reason. As we have shown, spectra obtained from both theories are essentially identical for relativistic neutrinos. Recent studies have suggested that flavor vacuum may be naturally realized when neutrino mixing is dynamically generated in chiral-symmetric models~\cite{Blasone:2018hah}.  Alternatively, one possible reason in favor of the mass eigenstates is that they represent elementary particle states whereas the coherent flavor eigenstates are not. To be specific, at finite-volume, the flavor vacuum is a condensate of particle-anti-particle mass eigenstate pairs~\cite{Blasone:1995zc}. If this is the case, it may be interesting to understand what are the conditions needed to produce coherent flavor eigenstates. At a foundational level, it is important to investigate the symmetries of the flavor vacuum. Existing results in the literature suggest that the flavor vacuum cannot be described by the irreducible representations Poincar\'{e}-group~\cite{Lobanov:2015esa,Blasone:2020cun}. In other words, if the flavor states are indeed physical, their observational signatures may constitute a direct evidence for Lorentz violation.
%For neutron $\beta$ decay, this means that either $n\rightarrow p^{+}+e^{-}+\overline{\nu}_{i}$ or
%$n\rightarrow p^{+}+e^{-}+\overline{\nu}_{\alpha}$ where $i=1,2,\cdots $ or $\alpha=e,\mu\cdots$ for the respective theories. 
%Lastly, there is in fact another possibility that involves entanglement. In this case, the neutrinos interact as mass eigenstates but they are entangled with the other decay products. For neutron $\beta$ decay, this means that the neutron state $|n\rangle$ decays to
%\begin{equation}
%|n\rangle\rightarrow\sum_{i}U_{ei}|\p^{(i)}_{\overline{\nu}}\rangle\otimes |\p^{(i)}_{p}\rangle\otimes|\p^{(i)}_{e}\rangle.
%\end{equation}
%To see how flavor oscillations arise, please see refs.~\cite{Cohen:2008qb,Ahluwalia:2011ea}.

%Finally, the theories that we have discussed may not be mutually exclusive since they all start with the relation $\nu_{\alpha}(x)=\sum_{i}U_{\alpha i}\nu_{i}(x)$. In particular, when neutrinos interact as mass eigenstates, the theory can produce  separable as well as entangled states. 
%The production of entangled states should make  further complications to the detection and production of neutrinos.
%Which process is favorable may be determined by comparing the decay rates. 
%It would be interesting to  compare the interactions and oscillations derived from these theories.

%----------------------------------------------------
\section*{Acknowledgements}
%----------------------------------------------------

I would like to thank Mohini Gupta, Prasanta Kumar Rath, Amit Roy and Santosh Roy for useful discussions; Dharam Vir Ahluwalia, Massimo Blasone and Kazuyuki Furuuchi for comments on the initial manuscript; Gaetano Luciano and Luca Smaldone for finding mistakes in the initial manuscript.

\appendix

%-----------------------------------------------------------------
\section{Flavor operators}\label{A}
%-----------------------------------------------------------------

We now derive $a_{\alpha}(\p,\sigma)$ and $b_{\alpha}(\p,\sigma)$ in terms of $a_{i}(\p,\sigma)$ and $b_{i}(\p,\sigma)$. The Dirac field $\nu_{i}(x)$ is normalized such that $u_{i}(\p,\sigma)$ satisfy (no summations over $i$)
\begin{eqnarray}
&& u_{i}^{\dag}(\p,\sigma)u_{i}(\p,\sigma')=2E_{i}\delta_{\sigma\sigma'},\\
&& u_{i}^{\dag}(\p,\sigma)v_{i}(-\p,\sigma')=0,\hspace{0.5cm}i=1,2.
\end{eqnarray}
Therefore, in the Heisenberg picture, the annihilation operators $a_{\alpha}(\p,\sigma)$ are given by
\begin{eqnarray}
&& a_{e}(\p,\sigma)=(2\pi)^{-3/2}\int \frac{d^{3}x}{\sqrt{2E_{1}}} e^{-i\mathbf{p\cdot x}}u_{1}^{\dag}(\p,\sigma)\nu_{e}(\x),\\\label{eq:alpha1}
&& a_{\mu}(\p,\sigma)=(2\pi)^{-3/2}\int \frac{d^{3}x}{\sqrt{2E_{2}}} e^{-i\mathbf{p\cdot x}}u_{2}^{\dag}(\p,\sigma)\nu_{\mu}(\x).
\end{eqnarray}
where $\nu_{\alpha}(\x)\equiv \nu_{\alpha}(0,\x)$. Similarly for $b_{e}(\p,\sigma)$, using the identities
\begin{eqnarray}
&& v_{i}^{\dag}(\p,\sigma)v_{i}(\p,\sigma')=2E_{i}\delta_{\sigma\sigma'},\\
&& v_{i}^{\dag}(\p,\sigma)u_{i}(-\p,\sigma')=0,\hspace{0.5cm}i=1,2
\end{eqnarray}
we obtain
\begin{eqnarray}
&& b_{e}(\p,\sigma)=(2\pi)^{-3/2}\int \frac{d^{3}x}{\sqrt{2E_{1}}} e^{i\mathbf{p\cdot x}}v_{1}^{\dag}(\p,\sigma)\nu_{e}(\x),\\
&& b_{\mu}(\p,\sigma)=(2\pi)^{-3/2}\int \frac{d^{3}x}{\sqrt{2E_{2}}} e^{i\mathbf{p\cdot x}}v_{2}^{\dag}(\p,\sigma)\nu_{\mu}(\x).
\end{eqnarray}
Substituting eqs.~(\ref{eq:cs1}) and (\ref{eq:cs2}) into the above expressions, the flavor operators are
\begin{eqnarray}
&&a_{e}(\p,\sigma)=a_{1}(\p,\sigma) \cos\theta% \nonumber\\
+\Big[U(\p,\sigma)a_{2}(\p,\sigma)+\sum_{\sigma'}V(\p,\sigma,\sigma')b_{2}^{\dag}(-\p,\sigma')\Big]\sin\theta, \label{eq:ae}\\
&&a_{\mu}(\p,\sigma)=a_{2}(\p,\sigma)\cos\theta
-\Big[U(\p,\sigma)a_{1}(\p,\sigma)-\sum_{\sigma'}V(\p,\sigma,\sigma')b_{1}^{\dag}(-\p,\sigma')\Big] \sin\theta, \label{eq:amu}\\
&&b_{e}(\p,\sigma)=b_{1}(\p,\sigma)\cos\theta 
+\Big[U(\p,\sigma)b_{2}(\p,\sigma)+\sum_{\sigma'}V(\p,\sigma,\sigma')a^{\dag}_{2}(-\p,\sigma')\Big]\sin\theta,\label{eq:bed}\\
&&b_{\mu}(\p,\sigma)= b_{2}(\p,\sigma)\cos\theta
-\Big[U(\p,\sigma)b_{1}(\p,\sigma)-\sum_{\sigma'}V(\p,\sigma,\sigma')a^{\dag}_{1}(-\p,\sigma')\Big] \sin\theta. \label{eq:bmud}
\end{eqnarray}
The coefficients $U(\p,\sigma)$ and $V(\p,\sigma,\sigma')$ are defined as 
\begin{eqnarray}
&& U(\p,\sigma)\equiv\frac{1}{\sqrt{4E_{1}E_{2}}}u^{\dag}_{1}(\p,\sigma)u_{2}(\p,\sigma),\label{eq:up}\\
&& V(\p,\sigma,\sigma')\equiv \frac{1}{\sqrt{4E_{1}E_{2}}}u^{\dag}_{1}(\p,\sigma)v_{2}(-\p,\sigma').\label{eq:vp}
\end{eqnarray}
They evaluate to
\begin{eqnarray}
&& U(\p,{\textstyle{\pm\frac{1}{2}}})=\frac{|\p|^{2}+(m_{1}+E_{1})(m_{2}+E_{2})}{\sqrt{4E_{1}E_{2}(m_{1}+E_{1})(m_{2}+E_{2})}}, \label{eq:u_12}\\
&& V(\p,{\textstyle{+\frac{1}{2}}},{\textstyle{+\frac{1}{2}}})=\frac{[(m_{1}+E_{1})-(m_{2}+E_{2})](p_{x}-ip_{y})}{\sqrt{4E_{1}E_{2}(m_{1}+E_{1})(m_{2}+E_{2})}},\\
&&V(\p,{\textstyle{+\frac{1}{2}}},{\textstyle{-\frac{1}{2}}})=\frac{[(m_{2}+E_{2})-(m_{1}+E_{1})]p_{z}}{\sqrt{4E_{1}E_{2}(m_{1}+E_{1})(m_{2}+E_{2})}}
\end{eqnarray}
and satisfy the following identities
\begin{align}
& V(\p,\textstyle{+\frac{1}{2},+\frac{1}{2}})=-V^{*}(\p,\textstyle{-\frac{1}{2},-\frac{1}{2}}), \label{eq:v1}\\
& V(\p,\textstyle{+\frac{1}{2},-\frac{1}{2}})=+V(\p,\textstyle{-\frac{1}{2},+\frac{1}{2}}),
\label{eq:v2}\\
&\sum_{\sigma'}V(\p,\sigma,\sigma')V^{*}(\p,\sigma,\sigma')=
\sum_{\sigma'}V(\p,\sigma',\sigma)V^{*}(\p,\sigma,\sigma')=1-U^{2}(\p,\sigma)\\
&\sum_{\sigma'}V(\p,\sigma,\sigma')V^{*}(\p,-\sigma,\sigma')=
\sum_{\sigma'}V(\p,\sigma',\sigma)V^{*}(\p,-\sigma,\sigma')=0.
\end{align}
For convenience, we define
\begin{equation}
V^{2}(\p,\sigma)\equiv\sum_{\sigma'}V(\p,\sigma,\sigma')V^{*}(\p,\sigma,\sigma').
\label{eq:v2p}
\end{equation}
Utilizing the symmetry $\nu_{\alpha}(\x)\leftrightarrow\nu_{i}(\x)$ when $\theta\rightarrow-\theta$, we obtain
\begin{align}
&a_{1}(\p,\sigma)=a_{e}(\p,\sigma) \cos\theta
-\Big[U(\p,\sigma)a_{\mu}(\p,\sigma)+\sum_{\sigma'}V(\p,\sigma,\sigma')b_{\mu}^{\dag}(-\p,\sigma')\Big]\sin\theta  \label{eq:a1}\\
&a_{2}(\p,\sigma)=a_{\mu}(\p,\sigma)\cos\theta
+\Big[U(\p,\sigma)a_{e}(\p,\sigma)-\sum_{\sigma'}V(\p,\sigma,\sigma')b_{e}^{\dag}(-\p,\sigma')\Big] \sin\theta \label{eq:a2}\\
&b_{1}(\p,\sigma)=b_{e}(\p,\sigma)\cos\theta 
-\Big[U(\p,\sigma)b_{\mu}(\p,\sigma)+\sum_{\sigma'}V(\p,\sigma,\sigma')a^{\dag}_{\mu}(-\p,\sigma')\Big]\sin\theta  \label{eq:b1}\\
&b_{2}(\p,\sigma)= b_{\mu}(\p,\sigma)\cos\theta
+\Big[U(\p,\sigma)b_{e}(\p,\sigma)-\sum_{\sigma'}V(\p,\sigma,\sigma')a^{\dag}_{e}(-\p,\sigma')\Big] \sin\theta \label{eq:b2}
\end{align}
At time $t$, the operators are given by
\begin{align}
&a_{e}(\p,\sigma,t)=a_{1}(\p,\sigma,t) \cos\theta
+\Big[U(\p,\sigma)a_{2}(\p,\sigma,t)+\sum_{\sigma'}V(\p,\sigma,\sigma')b_{2}^{\dag}(-\p,\sigma',t)\Big]\sin\theta  \label{eq:aet} \\
&a_{\mu}(\p,\sigma,t)=a_{2}(\p,\sigma,t)\cos\theta
-\Big[U(\p,\sigma)a_{1}(\p,\sigma,t)-\sum_{\sigma'}V(\p,\sigma,\sigma')b_{1}^{\dag}(-\p,\sigma',t)\Big] \sin\theta  \label{eq:amt}\\
&b_{e}(\p,\sigma,t)=b_{1}(\p,\sigma,t)\cos\theta 
+\Big[U(\p,\sigma)b_{2}(\p,\sigma,t)+\sum_{\sigma'}V(\p,\sigma,\sigma')a^{\dag}_{2}(-\p,\sigma',t)\Big]\sin\theta \label{eq:bet}\\
&b_{\mu}(\p,\sigma,t)= b_{2}(\p,\sigma,t)\cos\theta
-\Big[U(\p,\sigma)b_{1}(\p,\sigma,t)-\sum_{\sigma'}V(\p,\sigma,\sigma')a^{\dag}_{1}(-\p,\sigma',t)\Big] \sin\theta \label{eq:bmt}
\end{align}
where $a_{i}(\p,\sigma,t)=e^{-E_{i}t}a_{i}(\p,\sigma)$ and $b_{i}(\p,\sigma,t)=e^{-iE_{i}t}b_{i}(\p,\sigma)$.

The time-evolution of the flavor operators can be derived using 
\begin{eqnarray}
&&a_{i}(\p,\sigma,t)=e^{-E_{i}t}a_{i}(\p,\sigma) \label{eq:ait} \\ 
&& b_{i}(\p,\sigma,t)=e^{-E_{i}t}b_{i}(\p,\sigma) \label{eq:bit}.
\end{eqnarray} 
 The relations between flavor operators at time $t$ and $t'$ are obtained by substituting eqs.~(\ref{eq:ait}) and (\ref{eq:bit}) into (\ref{eq:aet}-\ref{eq:bmt}) and then express $a_{i}(\p,\sigma')$ and $b_{i}(\p,\sigma')$ in terms of flavor operators $a_{\alpha}(\p,\sigma)$ and $b_{\alpha}(\p,\sigma)$. The results are
\begin{eqnarray}
&& a_{\alpha}(\p,\sigma,t)=\sum_{\beta,\sigma'}\left[U_{\alpha\beta}(\p,\sigma,t)
a_{\beta}(\p,\sigma)+V_{\alpha\beta}(\p,\sigma,\sigma',t)
b^{\dag}_{\beta}(-\p,\sigma')\right]\label{eq:aat}\\
&& b_{\alpha}(\p,\sigma,t)=\sum_{\beta,\sigma'}\left[U_{\alpha\beta}(\p,\sigma,t)
b_{\beta}(\p,\sigma)+V_{\alpha\beta}(\p,\sigma,\sigma',t)
a^{\dag}_{\beta}(-\p,\sigma')\right]\label{eq:bat}
\end{eqnarray}
where
\begin{eqnarray}
&& U_{ee}(\p,\sigma,t)=e^{-iE_{1}t}\cos^{2}\theta+
\left[e^{-iE_{2}t}U^{2}(\p,\sigma)
+e^{iE_{2}t}V^{2}(\p,\sigma)\right]\sin^{2}\theta\\
&& U_{\mu\mu}(\p,\sigma,t)=e^{-iE_{2}t}\cos^{2}\theta+
\left[e^{-iE_{1}t}U^{2}(\p,\sigma)+e^{iE_{1}t}V^{2}(\p,\sigma)\right]\sin^{2}\theta\\
&& U_{e\mu}(\p,\sigma,t)=U_{\mu e}(\p,\sigma,t)=\left[e^{-iE_{2}t}-e^{-iE_{1}t}\right]\cos\theta\sin\theta \,U(\p,\sigma)\\
&& V_{ee}(\p,\sigma,\sigma',t)=\left[e^{iE_{2}t}-e^{-iE_{2}t}\right]
\sin^{2}\theta \,U(\p,\sigma)V(\p,\sigma,\sigma')\\
&& V_{\mu\mu}(\p,\sigma,\sigma',t)=\left[e^{-iE_{1}t}-e^{iE_{1}t}\right]\sin^{2}\theta\,U(\p,\sigma)V(\p,\sigma,\sigma')\\
&& V_{e\mu}(\p,\sigma,\sigma',t)=\left[e^{iE_{2}t}-e^{-iE_{1}t}\right]
\sin\theta\cos\theta \,V(\p,\sigma,\sigma')\\
&& V_{\mu e}(\p,\sigma,\sigma',t)=\left[e^{iE_{1}t}-e^{-iE_{2}t}\right]
\cos\theta\sin\theta\, V(\p,\sigma,\sigma')
\end{eqnarray}

We use eqs.~(\ref{eq:aat}) and (\ref{eq:bat}) to expand $\nu_{\alpha}(t,\x)$ in terms of flavor operators at time $t=0$. The results are
\begin{eqnarray}
&& \nu_{e}(t,\x)\equiv\lambda_{e}(t,\x)+\lambda_{\mu}(t,\x),\label{eq:net}\\
&& \nu_{\mu}(t,\x)\equiv\varrho_{\mu}(t,\x)+\varrho_{e}(t,\x)\label{eq:nmt}.
\end{eqnarray}
The field operators on the right-hand side are defined as
\begin{eqnarray}
\lambda_{\alpha}(t,\x)=(2\pi)^{-3/2}\int\frac{d^{3}p}{\sqrt{2E_{1}}}\sum_{\sigma,\sigma'}
\left[e^{i\mathbf{p\cdot x}}\,\mathcal{U}_{\alpha}(\p,\sigma,\sigma',t)a_{\alpha}(\p,\sigma')
+e^{-i\mathbf{p\cdot x}}\mathcal{V}_{\alpha}(\p,\sigma,\sigma',t)b^{\dag}_{\alpha}(\p,\sigma')\right]
\end{eqnarray}
\begin{eqnarray}
\varrho_{\alpha}(t,\x)=(2\pi)^{-3/2}\int\frac{d^{3}p}{\sqrt{2E_{2}}}\sum_{\sigma,\sigma'}
\left[e^{i\mathbf{p\cdot x}}\,\mathcal{X}_{\alpha}(\p,\sigma,\sigma',t)a_{\alpha}(\p,\sigma')
+e^{-i\mathbf{p\cdot x}}\mathcal{Y}_{\alpha}(\p,\sigma,\sigma',t)b^{\dag}_{\alpha}(\p,\sigma')\right]
\end{eqnarray}
The expansion coefficients are defined as
\begin{eqnarray}
&&\mathcal{U}_{e}(\p,\sigma,\sigma',t)=U_{ee}(\p,\sigma,t)\delta_{\sigma\sigma'}u_{1}(\p,\sigma)+V^{*}_{ee}(-\p,\sigma,\sigma',t)v_{1}(-\p,\sigma) \\
&&\mathcal{U}_{\mu}(\p,\sigma,\sigma',t)=U_{e\mu}(\p,\sigma,t)\delta_{\sigma\sigma'}u_{1}(\p,\sigma)+V^{*}_{e\mu}(-\p,\sigma,\sigma',t)v_{1}(-\p,\sigma) \\
&&\mathcal{V}_{e}(\p,\sigma,\sigma',t)=U^{*}_{ee}(\p,\sigma,t)\delta_{\sigma\sigma'}
v_{1}(\p,\sigma)+V_{ee}(-\p,\sigma,\sigma',t)u_{1}(-\p,\sigma) \label{eq:ve}\\
&&\mathcal{V}_{\mu}(\p,\sigma,\sigma',t)=U^{*}_{e\mu}(\p,\sigma,t)\delta_{\sigma\sigma'}v_{1}(\p,\sigma)+V_{e\mu}(-\p,\sigma,\sigma',t)u_{1}(-\p,\sigma) \label{eq:vm}\\
&&\mathcal{X}_{e}(\p,\sigma,\sigma',t)=U_{\mu e}(\p,\sigma,t)\delta_{\sigma\sigma'}u_{2}(\p,\sigma)+V^{*}_{ee}(-\p,\sigma,\sigma',t)v_{2}(-\p,\sigma)\\
&&\mathcal{X}_{\mu}(\p,\sigma,\sigma',t)=U_{\mu\mu}(\p,\sigma,t)\delta_{\sigma\sigma'}u_{2}(\p,\sigma)+V^{*}_{e\mu}(-\p,\sigma,\sigma',t)v_{2}(-\p,\sigma)\\
&&\mathcal{Y}_{e}(\p,\sigma,\sigma',t)=U^{*}_{\mu e}(\p,\sigma,t)\delta_{\sigma\sigma'}v_{2}(\p,\sigma)+V_{\mu e}(-\p,\sigma,\sigma',t)u_{2}(-\p,\sigma)\\
&&\mathcal{Y}_{\mu}(\p,\sigma,\sigma',t)=U^{*}_{\mu \mu}(\p,\sigma,t)\delta_{\sigma\sigma'}v_{2}(\p,\sigma)+V_{\mu e}(-\p,\sigma,\sigma',t)u_{2}(-\p,\sigma)
\end{eqnarray}

%---------------------------------------
\section{Transition rates}\label{tr}
%---------------------------------------

In a box of finite volume $V$ where the interaction is switched on for finite time $T$, the transition probability for $|A\rangle\rightarrow|B\rangle$ is~\cite{Weinberg:1995mt}
\begin{equation}
P(A\rightarrow B)=\left[\frac{(2\pi)^{3}}{V}\right]^{N_{A}+N_{B}}|S_{BA}|^{2}
\end{equation}
where $N_{A,B}$ is the number of particles in $|A\rangle$ and $|B\rangle$ and $S_{BA}$ is the $S$-matrix. The number of states in the phase space $dB=d^{3}p'_{1}\cdots d^{3}p'_{N_{B}}$ is
\begin{equation}
dN_{B}=\left[\frac{V}{(2\pi)^{3}}\right]^{N_{B}}dB.
\end{equation}
Therefore, the differential probability for $|A\rangle$ to end up in $|B\rangle$ within the range $dB$ is
\begin{equation}
dP(A\rightarrow B)=\left[\frac{(2\pi)^{3}}{V}\right]^{N_{A}}|S_{BA}|^{2}dB.
\end{equation}
In our calculation, the $S$-matrix contains more than one $\delta_{T}$ functions. For this reason, we factor the $S$-matrix as
\begin{equation}
S_{BA}\equiv-2\pi i \delta^{3}_{V}(\p_{B}-\p_{A})M_{BA}
\end{equation}
where
\begin{equation}
\delta^{3}_{V}(\p_{B}-\p_{A})=\frac{1}{(2\pi)^{3}}\int_{V}d^{3}x\, e^{i(\mathbf{p}_{B}-\mathbf{p}_{A})t}=\frac{V}{(2\pi)^{3}}\delta_{\mathbf{p}_{B}\mathbf{p}_{A}}
\end{equation}
and $\delta_{\mathbf{p}_{B}\mathbf{p}_{A}}$ is the Kronecker $\delta$ function.
Therefore, the square of $\delta^{3}_{V}(\p_{B}-\p_{A})$ is given by
\begin{equation}
\left[\delta^{3}_{V}(\p_{B}-\p_{A})\right]^{2}=\frac{V}{(2\pi)^{3}}\delta^{3}_{V}(\p_{B}-\p_{A}).
\end{equation}
The differential probability becomes
\begin{equation}
dP(A\rightarrow B)=(2\pi)^{2}\left[\frac{(2\pi)^{3}}{V}\right]^{N_{A}-1}|M_{BA}|^{2}
\delta^{3}_{V}(\p_{B}-\p_{A})dB.
\end{equation}
In the limit of large $V$, the $\delta^{3}_{V}(\p_{B}-\p_{A})$ becomes $\delta^{3}(\p_{B}-\p_{A})$. At large $T$, the energy conserving $\delta_{T}$ functions in $M_{BA}$ becomes the energy-conserving Dirac $\delta$ functions. Additionally, $|M_{BA}|^{2}$ will be proportional to $T$. The differential transition rate is defined as the differential probability per unit time
\begin{eqnarray}
d\Gamma(A\rightarrow B)&\equiv&\frac{dP(A\rightarrow B)}{T}\nonumber\\
&=&\frac{(2\pi)^{3N_{A}-1}V^{1-N_{A}}}{T}|M_{BA}|^{2}\delta^{3}(\p_{B}-\p_{A})dB
\end{eqnarray}

%---------------------------------------------------------
\section{Spin-sums and traces for neutron beta decay }
%---------------------------------------------------------

The spin-averaged decay rate for $n\rightarrow p^{+}+e^{-}+\overline{\nu}_{e}$ is\begin{eqnarray}
d\Gamma(\overline{\nu}_{e})&=&\frac{1}{2}\frac{(2\pi)^{2}}{T}\sum_{\tiny{\mbox{spins}}}|K^{\mu}M_{\mu}|^{2}\delta^{3}(\p_{p}+\p_{e}+\p_{\nu})d^{3}p_{p}d^{3}p_{e}d^{3}p_{\nu} \nonumber\\
&=&\frac{1}{2}\frac{(2\pi)^{2}}{T}\sum_{\tiny{\mbox{spins}}}\left[(K^{\mu}K^{\nu\dag})(M_{\mu}M^{\dag}_{\nu})\right]\delta^{3}(\p_{p}+\p_{e}+\p_{\nu})d^{3}p_{p}d^{3}p_{e}d^{3}p_{\nu}
\end{eqnarray}
The spin-sum for $M_{\mu}M^{\dag}_{\nu}$ is straightforward to evaluate
\begin{eqnarray}
\sum_{\tiny{\mbox{spins}}}M^{\mu}M^{\nu\dag}&=&
\frac{G^{2}_{F}V^{2}_{ud}}{(2\pi)^{8}(32E_{n}E_{p}E_{e}E_{1})}
\mbox{tr}\left[(\slashed{p}_{p}+m_{p})\gamma^{\mu}(f-\gamma^{5}g)
(\slashed{p}_{n}+m_{n})\gamma^{\nu}(f-\gamma^{5}g)\right]\nonumber\\
&=&\frac{G^{2}_{F}V^{2}_{ud}}{(2\pi)^{8}(16E_{n}E_{p}E_{e}E_{1})}\Big{[}
-2i\epsilon^{\mu\nu\rho\sigma}(p_{n})_{\rho}(p_{p})_{\sigma}fg+(f^{2}+g^{2})
(p^{\mu}_{n}p^{\nu}_{p}+p^{\mu}_{p}p^{\nu}_{n})\nonumber\\
&&\hspace{3.5cm}+\eta^{\mu\nu}[m_{n}m_{p}(f^{2}-g^{2})-(p_{n}\cdot p_{p})(f^{2}+g^{2})\Big{]}.
\end{eqnarray}
However, the spin-sum for $K^{\mu}K^{\nu\dag}$ is more complicated so we split it into three terms
\begin{equation}
\sum_{\tiny{\mbox{spins}}}K^{\mu}K^{\nu\dag}=\sum^{3}_{i=1}K^{\mu\nu}_{i}
\end{equation}
where
\begin{eqnarray}
K^{\mu\nu}_{1}&=&F^{2}\mbox{tr}\left[
(\slashed{p}_{e}+m_{e})\gamma^{\mu}(I-\gamma^{5})(\slashed{p}_{1}-m_{1})
\gamma^{\nu}(I-\gamma^{5})\right] \nonumber\\
&=&8F^{2}\left[-i\epsilon^{\mu\nu\rho\sigma}(p_{1})_{\rho}(p_{e})_{\sigma}
-\eta^{\mu\nu}(p_{1}\cdot p_{e})+(p^{\mu}_{1}p^{\nu}_{e}+p^{\nu}_{1}p^{\mu}_{e})\right]
\end{eqnarray}
and
\begin{eqnarray}
 K^{\mu\nu}_{2}&=&FGU\mbox{tr}\Big{\{}(\slashed{p}_{e}+m_{e})\gamma^{\mu}(I-\gamma^{5})\nonumber\\
&&\times \sum_{\sigma\sigma'\sigma_{\nu}}\Big{[}
 V^{*}(-\p_{\nu},\sigma',\sigma_{\nu})v_{1}(\p_{\nu},\sigma_{\nu})\overline{u}_{1}(-\p_{\nu},\sigma')+V(-\p_{\nu},\sigma,\sigma_{\nu})u_{1}(-\p_{\nu},\sigma)\overline{v}_{1}(\p_{\nu},\sigma_{\nu}\Big{]}\gamma^{\nu}(I-\gamma^{5})\Big{\}}\nonumber\\
\end{eqnarray}
\begin{align}
K^{\mu\nu}_{3}=&G^{2}U^{2}
\mbox{tr}\Big{[}(\slashed{p}_{e}+m_{e})\gamma^{\mu}(I-\gamma^{5})\nonumber\\
&\times\sum_{\sigma\sigma'\sigma_{\nu}}V(-\p_{\nu},\sigma,\sigma_{\nu})
V^{*}(-\p_{\nu},\sigma',\sigma_{\nu})u_{1}(-\p_{\nu},\sigma)\overline{u}_{1}(-\p_{\nu},\sigma')\gamma^{\nu}(I-\gamma^{5})\Big{]}
\end{align}
To evaluate the traces for $K^{\mu\nu}_{2}$ and $K^{\mu\nu}_{3}$, we rewrite the spin-sums using eqs.~(\ref{eq:v1}) and (\ref{eq:v2}). After some algebraic manipulations $K^{\mu\nu}_{2}$ becomes
\begin{eqnarray}
K^{\mu\nu}_{2}&=&FGU\mbox{tr}\Big{\{}(\slashed{p}_{e}+m_{e})\gamma^{\mu}(I-\gamma^{5})\nonumber\\
&&\times\Big{[}\sum_{\sigma_{\nu}}V(-\p_{\nu},\sigma_{\nu},\sigma_{\nu})
\left[u_{1}(-\p_{\nu},\sigma_{\nu})\overline{v}_{1}(\p_{\nu},\sigma_{\nu})
-v_{1}(\p_{\nu},-\sigma_{\nu})\overline{u}_{1}(-\p_{\nu},-\sigma_{\nu})\right]\nonumber\\
&&+\sum_{\sigma_{\nu}}V(-\p_{\nu},-\sigma_{\nu},\sigma_{\nu})
\left[u_{1}(-\p_{\nu},-\sigma_{\nu})\overline{v}_{1}(\p_{\nu},\sigma_{\nu})
+v_{1}(\p_{\nu},\sigma_{\nu})\overline{u}_{1}(-\p_{\nu},-\sigma_{\nu})\right]\Big{]}\gamma^{\nu}(I-\gamma^{5})\Big{\}}\nonumber\\
\end{eqnarray}
After some tedious but straightforward calculations, we obtain
\begin{eqnarray}
K^{\mu\nu}_{2}&=&FG
\frac{m_{1}[|\p_{\nu}|^{2}+(m_{1}+E_{1})(m_{2}+E_{2})](m_{1}-m_{2}+E_{1}-E_{2})}{2E_{1}E_{2}(m_{1}+E_{1})(m_{2}+E_{2})} \nonumber\\
&&\times\mbox{tr}\Big{\{}(\slashed{p}_{e}+m_{e})\gamma^{\mu}(I-\gamma^{5})
\left[|\p_{\nu}|^{2}+m_{1}(\slashed{p}_{1}-\gamma^{0}E_{1})\right]\gamma^{\nu}
(I-\gamma^{5})\Big{\}}\nonumber\\
&=&8FG
\frac{m_{1}[|\p_{\nu}|^{2}+(m_{1}+E_{1})(m_{2}+E_{2})](m_{1}-m_{2}+E_{1}-E_{2})}{2E_{1}E_{2}(m_{1}+E_{1})(m_{2}+E_{2})}\nonumber\\
&&\times\Big{\{}-i\epsilon^{\mu\nu\rho\sigma}(p_{1})_{\rho}(p_{e})_{\sigma}-\eta^{\mu\nu}(p_{1}\cdot p_{e})+(p^{\mu}_{1}p^{\nu}_{e}+p^{\nu}_{1}p^{\mu}_{e})\nonumber\\
&&-E_{1}\left[-i\epsilon^{0\mu\nu\rho}(p_{e})_{\rho}
-\eta^{\mu\nu}E_{e}+\eta^{\mu0}p^{\nu}_{e}+\eta^{\nu0}p^{\mu}_{e}\right]\Big{\}}
\end{eqnarray}
where $\widetilde{p}^{\,\mu}_{1}=(E_{1},-\p_{\nu})$. For $K^{\mu\nu}_{3}$, we obtain
\begin{eqnarray}
K^{\mu\nu}_{3}&=&2G^{2}U^{2}V^{2}\mbox{tr}
\left[(\slashed{p}_{e}+m_{e})\gamma^{\mu}(I-\gamma^{5})
(\widetilde{\slashed{p}}_{1}-m_{1})\gamma^{\nu}(I-\gamma^{5})\right]\nonumber\\
&=&16G^{2}U^{2}V^{2}\left[-i\epsilon^{\mu\nu\rho\sigma}(\widetilde{p}_{1})_{\rho}(p_{e})_{\sigma}
-\eta^{\mu\nu}(\widetilde{p}_{1}\cdot p_{e})+(\widetilde{p}^{\,\mu}_{1}p^{\nu}_{e}+\widetilde{p}^{\,\nu}_{1}p^{\mu}_{e})\right].
\end{eqnarray}
The contractions with $M^{\mu}$ are given by
\begin{eqnarray}
\sum_{\tiny{\mbox{spins}}}K^{\mu\nu}_{1}M_{\mu}M^{\dag}_{\nu}
&=&\frac{2G^{2}_{F}V^{2}_{ud}F^{2}}{(2\pi)^{8}(E_{n}E_{p}E_{e}E_{1})}
\Big{[}(f+g)^{2}(p_{n}\cdot p_{1})(p_{p}\cdot p_{e})+(f-g)^{2}(p_{n}\cdot p_{e})
(p_{p}\cdot p_{1})\nonumber\\
&&\hspace{3cm}-(f^{2}-g^{2})(m_{n}m_{p})(p_{e}\cdot p_{1})\Big{]}\label{eq:k1}
\end{eqnarray}
\begin{eqnarray}
\sum_{\tiny{\mbox{spins}}}K^{\mu\nu}_{2}M_{\mu}M^{\dag}_{\nu}
&=&\left[\frac{2G^{2}_{F}V^{2}_{ud}FG}{(2\pi)^{8}(E_{n}E_{p}E_{e}E_{1})}\right]
\frac{m_{1}\left[|\p_{\nu}|^{2}+(m_{1}+E_{1})(m_{2}+E_{2})\right](m_{1}-m_{2}+E_{1}-E_{2})}{2E_{1}E_{2}(m_{1}+E_{1})(m_{2}+E_{2})}\nonumber\\
&&\times\Big{[}(f+g)^{2}(p_{n}\cdot p_{1})(p_{p}\cdot p_{e})+(f-g)^{2}(p_{n}\cdot p_{e})
(p_{p}\cdot p_{1})
-(f^{2}-g^{2})(m_{n}m_{p})(p_{e}\cdot p_{1})\Big{]}\nonumber\\
&&-\left[\frac{G^{2}_{F}V^{2}_{ud}FG}{(2\pi)^{8}(E_{n}E_{p}E_{e})}\right]
\frac{m_{1}\left[|\p_{\nu}|^{2}+(m_{1}+E_{1})(m_{2}+E_{2})\right](m_{1}-m_{2}+E_{1}-E_{2})}{2E_{1}E_{2}(m_{1}+E_{1})(m_{2}+E_{2})}\nonumber\\
&&\times\Big{\{}-4[(p_{e}\cdot p_{n})E_{p}-(p_{e}\cdot p_{p})E_{n}]fg
-2(f^{2}+g^{2})(p_{n}\cdot p_{p})E_{e}\nonumber\\
&&-4E_{e}\left[m_{n}m_{p}(f^{2}-g^{2})-(p_{n}\cdot p_{p})(f^{2}+g^{2})\right]
+2(f^{2}+g^{2})\left[m_{n}(p_{p}\cdot p_{e})+(p_{n}\cdot p_{e})E_{p}\right] \nonumber\\
&&+2E_{e}[m_{n}m_{p}(f^{2}-g^{2})-(p_{n}\cdot p_{p})(f^{2}+g^{2})]\Big{\}}
\label{eq:k2}
\end{eqnarray}
\begin{eqnarray}
\sum_{\tiny{\mbox{spins}}}K^{\mu\nu}_{3}M_{\mu}M^{\dag}_{\nu}
&=&\frac{4G^{2}_{F}V^{2}_{ud}U^{2}V^{2}G^{2}}{(2\pi)^{8}(E_{n}E_{p}E_{e}E_{1})}
\Big{[}(f+g)^{2}(p_{n}\cdot \widetilde{p}_{1})(p_{p}\cdot p_{e})+(f-g)^{2}(p_{n}\cdot p_{e})
(p_{p}\cdot \widetilde{p}_{1})\nonumber\\
&&\hspace{3cm}-(f^{2}-g^{2})(m_{n}m_{p})(p_{e}\cdot \widetilde{p}_{1})\Big{]}
\label{eq:k3}
\end{eqnarray}

\label{Bibliography}
%\lhead{\emph{Bibliography}}  % Change the left side page header to "Bibliography"
\bibliographystyle{JHEP}  % Use the "unsrtnat" BibTeX style for formatting the Bibliography
\bibliography{Bibliography}  % The references (bibliography) information are stored in the file named "Bibliography.bib"

\providecommand{\href}[2]{#2}\begingroup\raggedright\begin{thebibliography}{10}

\bibitem{Cleveland:1998nv}
B.~T. Cleveland, T.~Daily, R.~Davis, Jr., J.~R. Distel, K.~Lande, C.~K. Lee,
  P.~S. Wildenhain, and J.~Ullman, {\it {Measurement of the solar electron
  neutrino flux with the Homestake chlorine detector}},  {\em Astrophys. J.}
  {\bf 496} (1998) 505--526.

\bibitem{Fukuda:1998mi}
{\bf Super-Kamiokande} Collaboration, Y.~Fukuda et~al., {\it {Evidence for
  oscillation of atmospheric neutrinos}},  {\em Phys. Rev. Lett.} {\bf 81}
  (1998) 1562--1567, [\href{http://xxx.lanl.gov/abs/hep-ex/9807003}{{\tt
  hep-ex/9807003}}].

\bibitem{Aguilar:2001ty}
{\bf LSND} Collaboration, A.~Aguilar-Arevalo et~al., {\it {Evidence for
  neutrino oscillations from the observation of anti-neutrino(electron)
  appearance in a anti-neutrino(muon) beam}},  {\em Phys. Rev.} {\bf D64}
  (2001) 112007, [\href{http://xxx.lanl.gov/abs/hep-ex/0104049}{{\tt
  hep-ex/0104049}}].

\bibitem{Ahmad:2002jz}
{\bf SNO} Collaboration, Q.~R. Ahmad et~al., {\it {Direct evidence for neutrino
  flavor transformation from neutral current interactions in the Sudbury
  Neutrino Observatory}},  {\em Phys. Rev. Lett.} {\bf 89} (2002) 011301,
  [\href{http://xxx.lanl.gov/abs/nucl-ex/0204008}{{\tt nucl-ex/0204008}}].

\bibitem{Aguilar:2007it}
{\bf MiniBooNE} Collaboration, A.~Aguilar-Arevalo et~al., {\it {A search for
  electron neutrino appearance at the $\Delta m^{2} \sim 1\mbox{eV}^{\,2}$
  scale}},  {\em Phys. Rev. Lett.} {\bf 98} (2007) 231801,
  [\href{http://xxx.lanl.gov/abs/0704.1500}{{\tt arXiv:0704.1500}}].

\bibitem{Adamson:2011qu}
{\bf MINOS} Collaboration, P.~Adamson et~al., {\it {Improved search for
  muon-neutrino to electron-neutrino oscillations in MINOS}},  {\em Phys. Rev.
  Lett.} {\bf 107} (2011) 181802,
  [\href{http://xxx.lanl.gov/abs/1108.0015}{{\tt arXiv:1108.0015}}].

\bibitem{Abe:2011fz}
{\bf Double Chooz} Collaboration, Y.~Abe et~al., {\it {Indication of reactor
  $\bar{\nu}_e$ disappearance in the Double Chooz experiment}},  {\em Phys.
  Rev. Lett.} {\bf 108} (2012) 131801,
  [\href{http://xxx.lanl.gov/abs/1112.6353}{{\tt arXiv:1112.6353}}].

\bibitem{An:2012eh}
{\bf Daya Bay} Collaboration, F.~P. An et~al., {\it {Observation of
  electron-antineutrino disappearance at Daya Bay}},  {\em Phys. Rev. Lett.}
  {\bf 108} (2012) 171803, [\href{http://xxx.lanl.gov/abs/1203.1669}{{\tt
  arXiv:1203.1669}}].

\bibitem{Ahn:2012nd}
{\bf RENO} Collaboration, J.~K. Ahn et~al., {\it {Observation of reactor
  electron antineutrino Disappearance in the RENO Experiment}},  {\em Phys.
  Rev. Lett.} {\bf 108} (2012) 191802,
  [\href{http://xxx.lanl.gov/abs/1204.0626}{{\tt arXiv:1204.0626}}].

\bibitem{Shrock:1980vy}
R.~E. Shrock, {\it {New tests for, and bounds on, neutrino masses and lepton
  mixing}},  {\em Phys. Lett.} {\bf 96B} (1980) 159--164.

\bibitem{Shrock:1980ct}
R.~E. Shrock, {\it {General theory of weak leptonic and semileptonic decays. 1.
  Leptonic pseudoscalar meson decays, with associated tests for, and bounds on,
  neutrino masses and lepton Mixing}},  {\em Phys. Rev.} {\bf D24} (1981) 1232.

\bibitem{Shrock:1981wq}
R.~E. Shrock, {\it {General theory of weak processes involving neutrinos. 2.
  Pure Leptonic Decays}},  {\em Phys. Rev.} {\bf D24} (1981) 1275.

\bibitem{Giunti:1991cb}
C.~Giunti, C.~W. Kim, and U.~W. Lee, {\it {Comments on the weak states of
  neutrinos}},  {\em Phys. Rev.} {\bf D45} (1992) 2414--2420.

\bibitem{Giunti:1993se}
C.~Giunti, C.~W. Kim, J.~A. Lee, and U.~W. Lee, {\it {On the treatment of
  neutrino oscillations without resort to weak eigenstates}},  {\em Phys. Rev.}
  {\bf D48} (1993) 4310--4317,
  [\href{http://xxx.lanl.gov/abs/hep-ph/9305276}{{\tt hep-ph/9305276}}].

\bibitem{Blasone:1995zc}
M.~Blasone and G.~Vitiello, {\it {Quantum field theory of fermion mixing}},
  {\em Annals Phys.} {\bf 244} (1995) 283--311,
  [\href{http://xxx.lanl.gov/abs/hep-ph/9501263}{{\tt hep-ph/9501263}}].
  [Erratum: Annals Phys.249,363(1996)].

\bibitem{Li:2006qt}
Y.~F. Li and Q.~Y. Liu, {\it {A Paradox on quantum field theory of neutrino
  mixing and oscillations}},  {\em JHEP} {\bf 10} (2006) 048,
  [\href{http://xxx.lanl.gov/abs/hep-ph/0604069}{{\tt hep-ph/0604069}}].

\bibitem{Blasone:2006jx}
M.~Blasone, A.~Capolupo, C.-R. Ji, and G.~Vitiello, {\it {On flavor
  conservation in weak interaction decays involving mixed neutrinos}},  {\em
  Int. J. Mod. Phys.} {\bf A25} (2010) 4179--4194,
  [\href{http://xxx.lanl.gov/abs/hep-ph/0611106}{{\tt hep-ph/0611106}}].

\bibitem{Osipowicz:2001sq}
{\bf KATRIN} Collaboration, A.~Osipowicz et~al., {\it {KATRIN: A Next
  generation tritium beta decay experiment with sub-eV sensitivity for the
  electron neutrino mass. Letter of intent}},
  \href{http://xxx.lanl.gov/abs/hep-ex/0109033}{{\tt hep-ex/0109033}}.

\bibitem{WEINHEIMER2002141}
C.~Weinheimer, {\it $\mbox{KATRIN}$, a next generation tritium $\beta$ decay
  experiment in search for the absolute neutrino mass scale},  {\em Progress in
  Particle and Nuclear Physics} {\bf 48} (2002), no.~1 141 -- 150.

\bibitem{Betts:2013uya}
S.~Betts et~al., {\it {Development of a relic neutrino detection experiment at
  PTOLEMY: Princeton Tritium Observatory for Light, Early-Universe,
  Massive-Neutrino Yield}},  in {\em {Proceedings, 2013 community summer study
  on the future of U.S. particle physics: Snowmass on the Mississippi
  (CSS2013): Minneapolis, MN, USA, July 29-August 6, 2013}}, 2013.
\newblock \href{http://xxx.lanl.gov/abs/1307.4738}{{\tt arXiv:1307.4738}}.

\bibitem{Chang:1980qw}
L.-N. Chang and N.-P. Chang, {\it {Structure of the vacuum and neutron and
  neutrino Oscillations}},  {\em Phys. Rev. Lett.} {\bf 45} (1980) 1540.

\bibitem{Blasone:2002jv}
M.~Blasone, A.~Capolupo, and G.~Vitiello, {\it {Quantum field theory of three
  flavor neutrino mixing and oscillations with CP violation}},  {\em Phys.
  Rev.} {\bf D66} (2002) 025033,
  [\href{http://xxx.lanl.gov/abs/hep-th/0204184}{{\tt hep-th/0204184}}].

\bibitem{Hannabuss:2000hy}
K.~C. Hannabuss and D.~C. Latimer, {\it {The quantum field theory of fermion
  mixing}},  {\em J. Phys.} {\bf A33} (2000) 1369--1373.

\bibitem{Ji:2002tx}
C.-R. Ji and Y.~Mishchenko, {\it {The General theory of quantum field mixing}},
   {\em Phys. Rev.} {\bf D65} (2002) 096015,
  [\href{http://xxx.lanl.gov/abs/hep-ph/0201188}{{\tt hep-ph/0201188}}].

\bibitem{Blasone:2003eh}
M.~Blasone and G.~Vitiello, {\it {Quantum field theory of particle mixing and
  oscillations}},  in {\em {8th International Symposium Symmetries in
  Science}}, 9, 2003.
\newblock \href{http://xxx.lanl.gov/abs/hep-ph/0309202}{{\tt hep-ph/0309202}}.

\bibitem{Fujii:1998xa}
K.~Fujii, C.~Habe, and T.~Yabuki, {\it {Note on the field theory of neutrino
  mixing}},  {\em Phys. Rev.} {\bf D59} (1999) 113003,
  [\href{http://xxx.lanl.gov/abs/hep-ph/9807266}{{\tt hep-ph/9807266}}].
  [Erratum: Phys. Rev.D60,099903(1999)].

\bibitem{Fujii:2001zv}
K.~Fujii, C.~Habe, and T.~Yabuki, {\it {Remarks on flavor neutrino propagators
  and oscillation formulae}},  {\em Phys. Rev.} {\bf D64} (2001) 013011,
  [\href{http://xxx.lanl.gov/abs/hep-ph/0102001}{{\tt hep-ph/0102001}}].

\bibitem{Giunti:2003dg}
C.~Giunti, {\it {Fock states of flavor neutrinos are unphysical}},  {\em Eur.
  Phys. J. C} {\bf 39} (2005) 377--382,
  [\href{http://xxx.lanl.gov/abs/hep-ph/0312256}{{\tt hep-ph/0312256}}].

\bibitem{Blasone:2005ae}
M.~Blasone, A.~Capolupo, F.~Terranova, and G.~Vitiello, {\it {Lepton charge and
  neutrino mixing in pion decay processes}},  {\em Phys. Rev. D} {\bf 72}
  (2005) 013003, [\href{http://xxx.lanl.gov/abs/hep-ph/0505178}{{\tt
  hep-ph/0505178}}].

\bibitem{Blasone:2010zn}
M.~Blasone, M.~Di~Mauro, and G.~Vitiello, {\it {Non-abelian gauge structure in
  neutrino mixing}},  {\em Phys. Lett. B} {\bf 697} (2011) 238--245,
  [\href{http://xxx.lanl.gov/abs/1003.5812}{{\tt arXiv:1003.5812}}].

\bibitem{Giunti:2007ry}
C.~Giunti and C.~W. Kim, {\it {Fundamentals of neutrino physics and
  astrophysics}},  {\em Oxford, UK: Univ. Pr.} (2007) 710 p.

\bibitem{Giunti:1997wq}
C.~Giunti and C.~W. Kim, {\it {Coherence of neutrino oscillations in the wave
  packet approach}},  {\em Phys. Rev.} {\bf D58} (1998) 017301,
  [\href{http://xxx.lanl.gov/abs/hep-ph/9711363}{{\tt hep-ph/9711363}}].

\bibitem{Giunti:2002xg}
C.~Giunti, {\it {Neutrino wave packets in quantum field theory}},  {\em JHEP}
  {\bf 11} (2002) 017, [\href{http://xxx.lanl.gov/abs/hep-ph/0205014}{{\tt
  hep-ph/0205014}}].

\bibitem{Blasone:2002wp}
M.~Blasone, P.~Pires~Pacheco, and H.~Wan Chan~Tseung, {\it {Neutrino
  oscillations from relativistic flavor currents}},  {\em Phys. Rev.} {\bf D67}
  (2003) 073011, [\href{http://xxx.lanl.gov/abs/hep-ph/0212402}{{\tt
  hep-ph/0212402}}].

\bibitem{Schiavilla:1998je}
R.~Schiavilla et~al., {\it {Weak capture of protons by protons}},  {\em Phys.
  Rev.} {\bf C58} (1998) 1263,
  [\href{http://xxx.lanl.gov/abs/nucl-th/9808010}{{\tt nucl-th/9808010}}].

\bibitem{Long:2014zva}
A.~J. Long, C.~Lunardini, and E.~Sabancilar, {\it {Detecting non-relativistic
  cosmic neutrinos by capture on Tritium: phenomenology and physics
  potential}},  {\em JCAP} {\bf 1408} (2014) 038,
  [\href{http://xxx.lanl.gov/abs/1405.7654}{{\tt arXiv:1405.7654}}].

\bibitem{Blasone:2018hah}
M.~Blasone, P.~Jizba, N.~E. Mavromatos, and L.~Smaldone, {\it {Dynamical
  generation of field mixing via flavor vacuum condensate}},  {\em Phys. Rev.
  D} {\bf 100} (2019), no.~4 045027,
  [\href{http://xxx.lanl.gov/abs/1807.0761}{{\tt arXiv:1807.0761}}].

\bibitem{Lobanov:2015esa}
A.~Lobanov, {\it {Particle quantum states with indefinite mass and neutrino
  oscillations}},  {\em Annals Phys.} {\bf 403} (2019) 82--105,
  [\href{http://xxx.lanl.gov/abs/1507.0125}{{\tt arXiv:1507.0125}}].

\bibitem{Blasone:2020cun}
M.~Blasone, P.~Jizba, N.~E. Mavromatos, and L.~Smaldone, {\it {Some nontrivial
  aspects of Poincar\'e and CPT invariance of flavor vacuum}},  {\em Phys. Rev.
  D} {\bf 102} (2020), no.~2 025021,
  [\href{http://xxx.lanl.gov/abs/2002.1107}{{\tt arXiv:2002.1107}}].

\bibitem{Weinberg:1995mt}
S.~Weinberg, {\it {The quantum theory of fields. Vol. 1: Foundations}},  {\em
  Cambridge, UK: Univ. Pr.} (1995) 609 p.

\end{thebibliography}\endgroup
\end{document}